\documentclass[journal]{IEEEtran}

\pdfoutput=1
\usepackage{mathrsfs}
\usepackage{amsfonts,amssymb} 
\usepackage{amsmath,amssymb,amsthm,lipsum}
\usepackage{algorithm}
\usepackage{algorithmic}
\floatname{algorithm}{Network}
\usepackage{graphicx}
\usepackage{bm}
\usepackage{CJK}
\usepackage{enumerate}
\usepackage{diagbox}
\usepackage{float}
\usepackage{caption}
\usepackage{cite}
\usepackage{bbm}
\usepackage{subfigure}
\usepackage{enumitem}
\usepackage{setspace}
\usepackage{bbding}
\usepackage{pifont}
\usepackage{wasysym}
\usepackage{tablefootnote}
\ifCLASSINFOpdf
\else
\fi
\begin{document}
	\title{Downlink Massive MIMO Channel Estimation via Deep Unrolling : Sparsity Exploitations in Angular Domain}

\author{An~Chen,~\IEEEmembership{Student Member,~IEEE,}
        Wenbo Xu,~\IEEEmembership{Member,~IEEE,}
		Liyang~Lu, ~\IEEEmembership{Student Member,~IEEE,}
        and~Yue Wang,~\IEEEmembership{Senior Member,~IEEE}
\thanks{A. Chen, W. Xu and L. Lu are with the Key Lab of Universal Wireless Communications, Ministry of Education, Beijing University of Posts and Telecommunications.

Y. Wang is with the Department of Electrical and Computer Engineering, George Mason University, Fairfax, VA.

W. Xu (e-mail: xuwb@bupt.edu.cn) and Y. Wang (e-mail: ywang56@gmu.edu) are the corresponding authors.}}

\maketitle

\begin{abstract}
In frequency division duplex (FDD) massive MIMO systems, reliable downlink channel estimation is essential for the subsequent data transmission but is realized at the cost of massive pilot overhead due to hundreds of antennas at base station (BS). In order to reduce the pilot overhead without compromising the estimation, compressive sensing (CS) based methods have been widely applied for channel estimation by exploiting the inherent sparse structure of massive MIMO channel in angular domain. However, they still suffer from high complexity during optimization process and the requirement of prior knowledge on sparsity information. To overcome these challenges, this paper develops a novel hybrid channel estimation framework by integrating the model-driven CS and data-driven deep unrolling techniques. The proposed framework is composed of a coarse estimation part and a fine correction part, which is implemented in a two-stage manner to exploit both inter- and intra-frame sparsities of channels in angular domain. Then, two estimation schemes are designed depending on whether priori sparsity information is required, where the second scheme designs a new thresholding function to eliminate such requirement. Numerical results are provided to verify that our schemes can achieve high accuracy with low pilot overhead and low complexity.
\end{abstract}
 

\IEEEpeerreviewmaketitle

\section{Introduction}

\IEEEPARstart{M}{assive MIMO} employing hundreds of antennas boosts the transmission capacity and spectral efficiency via large spatial multiplexing gain. The effects of additive noise and rayleigh fading in this system become negligible and high data rate can be achieved as the number of BS antennas grows to infinity \cite{1}. Such merits meet the fast-growing requirements of future wireless systems for huge capacity and wide coverage, which enable massive MIMO to be an attractive enabling technique in wireless communication \cite{58}. 

In massive MIMO systems, one important step is to accurately estimate channel state information (CSI) to facilitate precoding and beamforming for the subsequent data transmissions. Traditional channel estimation methods can be classified into two categories: The first is the blind-based methods which need short or no pilot sequences to perform channel estimation\cite{34,37,39,40}. However, they require prior knowledge in terms of the statistical information of the unknown channel\cite{39,40} or are realized over multiple data blocks \cite{34,37} and thus suffer from slow convergence, which limits their practicability. The second is the pilot-based methods \cite{5,41},  which almost require no prior information about the channel and the estimation is simple to implement. Therefore, the pilot-based methods are adopted in this paper to enable massive MIMO downlink transmission. However, the length of required pilot sequences increases proportionally with the number of antennas at BS,  which involves a large amount of channel coefficients to estimate and causes unaffordable pilot overhead cost \cite{2}.

In order to efficiently obtain accurate CSI estimation while reducing the pilot overhead, CS-based methods have been widely studied as an important candidate technique by exploiting the inherent sparsity of MIMO channels. Among different types of sparse channel structures, sparsity in angular domain is a typical one that is often studied and utilized in CS-based MIMO channel estimation methods, e.g.,  \cite{8,9,10,22,33,43}. Such sparsity stems from the fact that millimeter-wave (mmWave) propagation experiences limited scattering at BS with sparse multipaths, which induces limited angular directionalities of mmWave massive MIMO channels \cite{43,53,57}.

In terms of sparsity in angular-domain, both the intra- and inter-frame sparsity are exploited to improve the massive MIMO channel estimation performance \cite{9,33,10,22,44,48,56}, where the latter is further divided into small-scale one and large-scale one. Firstly, the intra-frame sparsity is caused by rich local scatterings at the UE side and limited scatterings at the BS side, where the spatial multipaths depart from BS in limited angles of depatures (AoDs) and arrive at UE in rich isotropic angles of arrivals (AoAs) \cite{22,49,50}. Therefore, within one frame, the channel in angular domain exhibits sparsity as shown in \cite{9,10,22,33,44,48}.
Secondly, the inter-frame sparsity is measured in small-scale \cite{9,33} or large-scale \cite{10,22,44,48,56}, which is introduced by the fact that some common AoDs are shared when UE moves slowly and receives downlink frames at physically closed locations \cite{9,49,50}.
The small-scale inter-frame sparsity is induced by a large number of shared AoDs between two adjacent frames, while the large-scale inter-frame sparsity is induced by a small number of shared AoDs among multiple successive frames. The aforementioned CS-based methods only consider one kind of sparsity with regard to the inter-frame sparsity, which leaves great space for improving the channel estimation performance. 

For traditional methods, both the pilot overhead due to large antenna arrays and the time complexity for reconstruction optimization are high. For these issues, deep unrolling technique has appeared as an effective solution. Recently, some CS-based reconstruction algorithms are unrolled into deep learning networks. Such deep unrolled networks are proposed since they can utilize the advantages of available large-volume data to learn the latent sparse structure while reduce the number of iterations required by conventional convex optimization algorithms. For example, based on the learned iterative soft thresholding algorithm (LISTA) \cite{18}, a series of deep unrolled networks have been developed, such as LISTA-coupling weight and support selection (LISTA-CPSS) \cite{19}, and LISTA-group sparse (LISTA-GS) \cite{20}, by incorporating prior information to refine the thresholding function and gradient descent step. Meanwhile, such deep unrolling techniques are also successfully introduced into the channel estimation area. An end-to-end structure is developed by unifying a pilot design network with an unrolled channel estimation network to improve channel estimation performance \cite{16}. By considering the channel estimation error in real scenario, the network unrolling is conducted iteratively between channel estimation and signal detection in \cite{31}, where channel estimation error is reduced by detected signals. Although the CS-based deep unrolled networks can improve the channel estimation performance, the exploitations of both intra- and inter-frame sparsities of massive MIMO channels are often ignored by the current literatures, which can help further reduce the pilot overhead without compromising the performance.

In this paper, a two-stage CS-based deep unrolling structure is proposed, by exploiting both of the intra- and inter-frame sparsities in massive MIMO channel. Firstly, we propose to jointly exploit the large-scale and small-scale inter-frame sparsities of sparse channels in multiple frames, which is generally ignored by the current literatures and can further reduce the pilot overhead compared with those schemes that exploit either kind of sparsity\cite{9,10,22}.  Moreover, to eliminate the impractical requirement of sparsity bound information in \cite{9,10,22}, a new scheme is proposed by utilizing the `first significant jump' rule \cite{26}.  Finally, the high complexity in the traditional optimization-based schemes \cite{9,10,22} is relieved via the deep unrolling technique by reducing the number of iterations. The main contributions of this paper are summarized as follows:

\begin{itemize}[leftmargin=1pt, labelsep=6pt,itemindent=10pt]

\item \textbf{Two-stage channel estimation structure:}
This structure consists of a coarse estimation net and a fine correction net by utilizing deep unrolling technique. The former net aims to coarsely estimate the channel by exploiting the large-scale inter-frame sparsity invoked by the shared AoDs among multiple successive frames. The latter net refines the coarse channel estimation frame by frame and exploits the small-scale inter-frame sparsity between two adjacent frames. To the best of our knowledge, this is the first work to utilize both of these two sparsities and combine them with deep unrolling techniques. 

\item \textbf{A novel scheme without the requirement of prior knowledge:} 
With the purpose to eliminate the requirement of the priori sparsity bound information, we propose a new scheme which is named coarse estimation net and fine correction net with block `first significant jump' thresholding function (C-F-BFSJ). To better present this scheme, we also propose a straightforward but naive scheme that requires the priori sparsity bound information, i.e., coarse estimation net and fine correction net with block support selection function (C-F-BSS), which generalizes support selection thresholding function in \cite{19}. The priori sparsity bound information is used in BSS thresholding function to decide the indices of the nonzero entries, but it is eliminated in BFSJ funtion with the utilization of `first significant jump'.

\item \textbf{Theoretical analyses about the parameter setup and computational complexity:} 
Firstly, we derive the formula of a parameter that indicates the average number of row-sparsity in the concatenated channel matrix. This parameter is a significant thresholding parameter utilized in the proposed BSS function to control the nonzero thresholding process. A novel theoretical analysis on the inference of this parameter is also developed.  Secondly, detailed computational complexity analyses of the proposed C-F-BFSJ and C-F-BSS are provided, where we show the advantage of our schemes in decreasing complexity compared with the schemes without applying deep unrolling technique.
\end{itemize}

The rest of this paper is organized as follows: In Section ${\rm\uppercase\expandafter{\romannumeral2}}$, the system model and the two-stage structure are described. In Section ${\rm\uppercase\expandafter{\romannumeral3}}$, we present two different schemes based on the proposed structure to solve the channel estimation problem. In Section ${\rm\uppercase\expandafter{\romannumeral4}}$, simulation results are presented to indicate the superiority of the proposed schemes over the traditional schemes, followed by conclusions in Section ${\rm\uppercase\expandafter{\romannumeral5}}$.

$Notations$: Lowercase and uppercase boldface letters stand for vectors and matrices (e.g., $\mathbf{a}$ and $\mathbf{A}$), respectively. For a matrix $\mathbf{A}$, we denote its transpose, conjugate transpose and trace by $\mathbf{A}^T$, $\mathbf{A}^H$ and $Tr(\mathbf{A})$, respectively. $\mathbf{0}_{M\times{N}}$ denotes a zero matrix with $M$ rows and $N$ columns. The entry in the $k$-th row and $l$-th column of a matrix is represented as $\mathbf{A}_{(k,l)}$. The $k$-th row and $l$-th column of a matrix are denoted as $\mathbf{A}_{(k,:)}$ and $\mathbf{A}_{(:,l)}$, respectively. $\mathbf{h}_{(k)}$ stands for the $k$-th entry in a vector. $supp(\mathbf{h})=\{j: \mathbf{h}_{(j)} \neq 0\}$ means the support set of a sparse vector $\mathbf{h}$. The $l_2$ norm, $l_{2,1}$ norm and the Frobenius norm are given by $||\cdot||_2$, $||\mathbf{G}||_{2,1}=\sum_{j=1}^M ||\mathbf{G}_{(j,:)}||_2$ and $||\cdot||_F$, respectively. $\mathbbm{R}(\cdot)$ and $\mathbbm{I}(\cdot)$ represent the real part and imaginary part of a complex matrix, respectively. $|\mathbf{\Gamma}|$ means the cardinality of the set $\mathbf{\Gamma}$. $\mathbf{\Gamma} \backslash \mathbf{\Lambda}$ denotes the set composed of elements in $\mathbf{\Gamma}$ while not in $\mathbf{\Lambda}$.  $\lfloor x \rfloor$ denotes the floor function that outputs the greatest integer less than or equal to $x$. $\mathbbm{E}(\cdot)$ denotes the mathematic expectation of a random variable. 

\section{SYSTEM MODEL AND PROBLEM DESCRIPTIONS}

\subsection{Channel Model}

\begin{figure}
  \centering
  \includegraphics[scale=0.50]{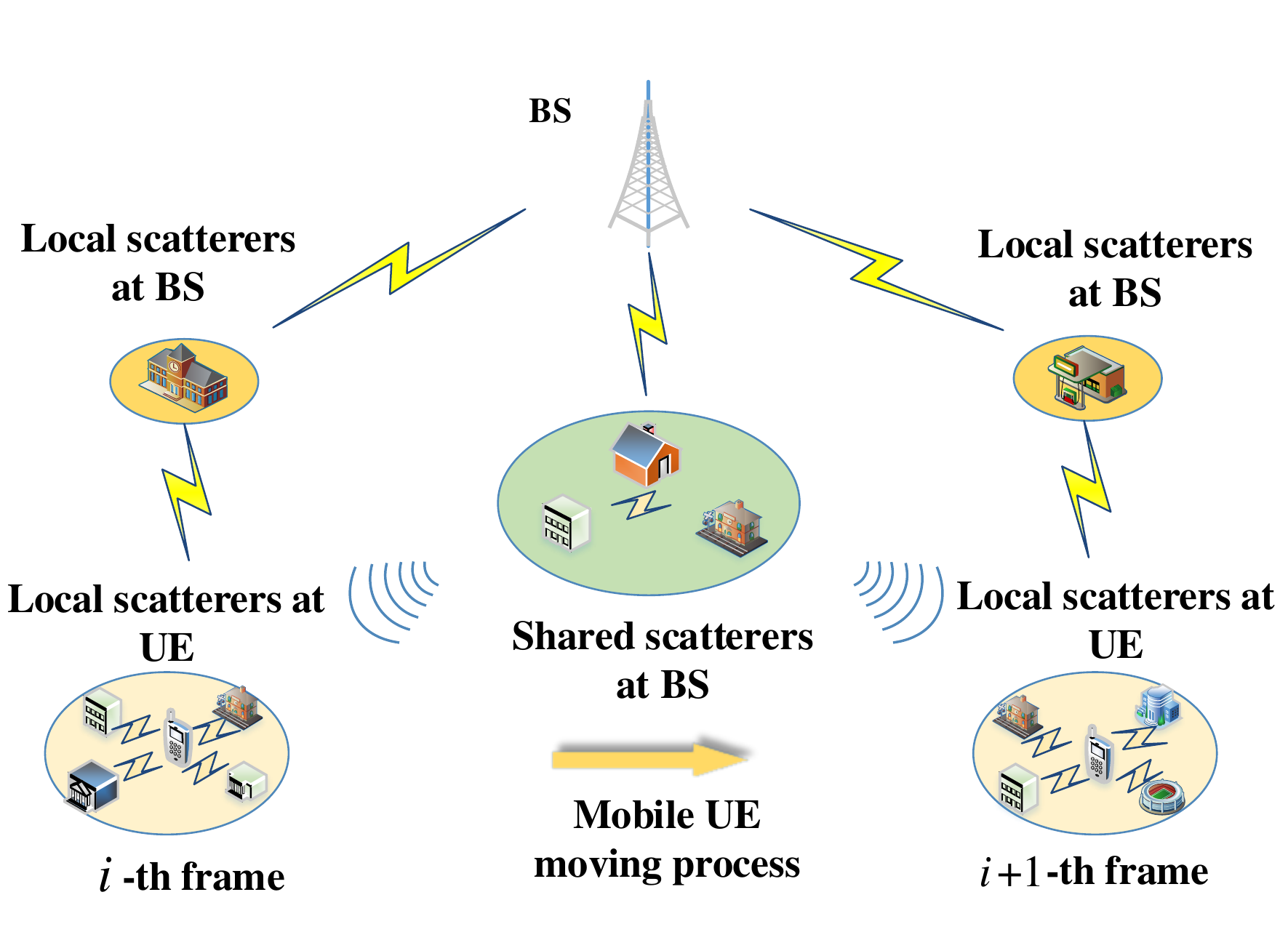}\\
  \caption{Illustration of the downlink MIMO system.}\label{system}
\end{figure}

In this paper, we consider FDD downlink massive MIMO system that is depicted in Fig. \ref{system}, which consists of one BS with $M$ transmit antennas and one mobile UE with $N$ receive antennas ($N \ll M$)$\footnote{The solutions developed in this work can be extended to multi-user massive MIMO systems as well.}$. Due to the mobility of UE, the downlink signals are influenced by rich local scatterers at UE side and limited scatterers at BS side. Those limited scatterers at BS include the shared scatterers that simultaneously affect the channel among consecutive frames, and the individual scatterers that only affect the channel in the current frame. To estimate the channel, BS sends pilot signals of length $T$ in each frame for each antenna and the received signals in the $i$-th frame can be expressed as \cite{9,10,22,43}:
\begin{equation}\label{CSmodel}
\mathbf{Y}^{[i]}= \mathbf{H}^{[i]}\mathbf{X} + \mathbf{N}^{[i]}, 
\end{equation}
where $\mathbf{H}^{[i]}\in \mathbb{C}^{N\times M}$ represents the downlink channel from BS to UE, $\mathbf{X}\in \mathbb{C}^{M\times T}$ denotes the transmitted pilot matrix, $Tr(\mathbf{X}^H\mathbf{X})=T$, 
and $\mathbf{N}^{[i]}$ is the noise matrix whose elements are i.i.d. complex Gaussian random variables with zero mean and variance $\delta^2$.

\begin{figure}
  \centering
  \includegraphics[scale=0.52]{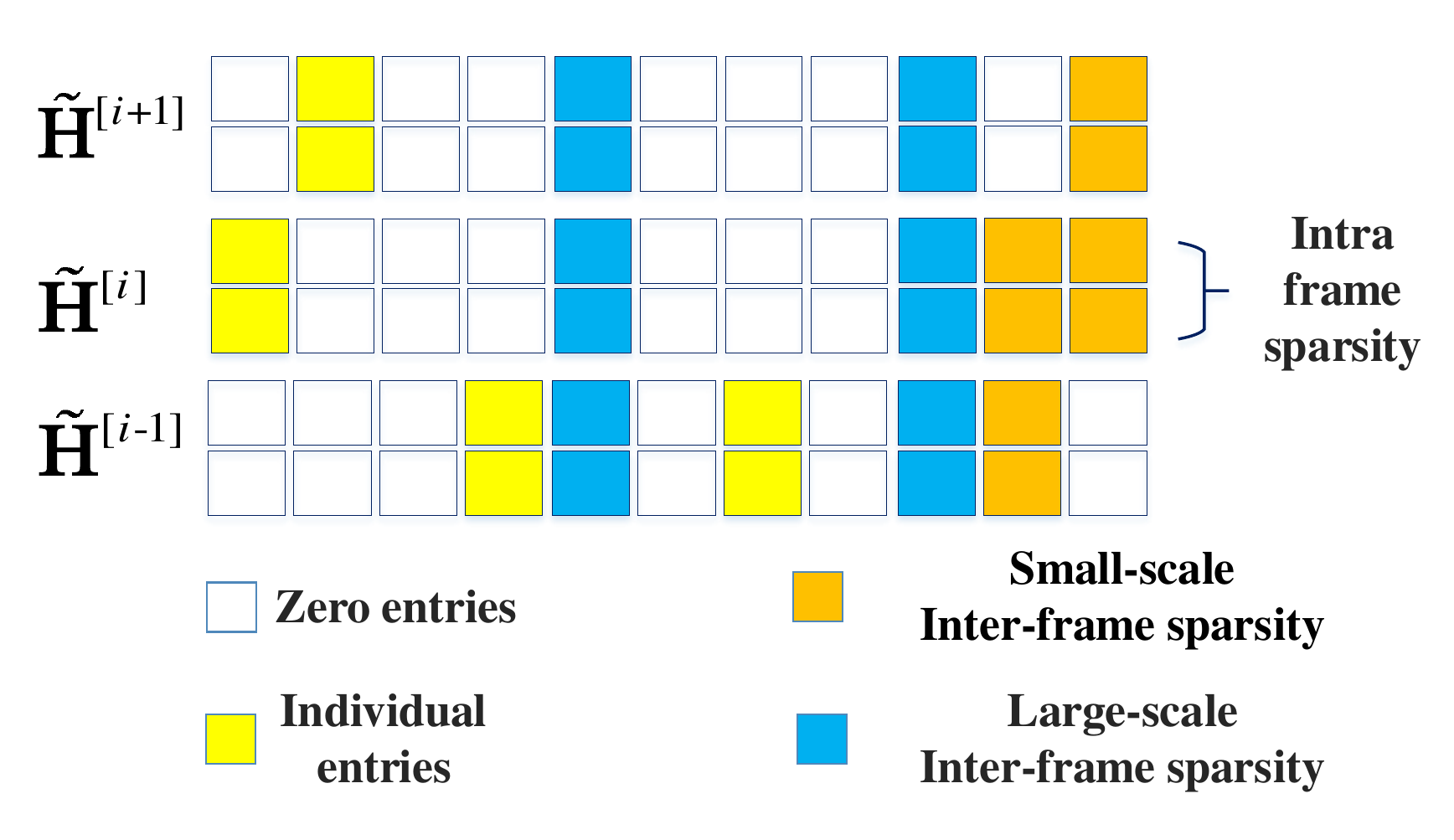}\\
  \caption{Illustration of angular domain channel model. Different colours in this figure represent different kinds of spatial paths affected by different scatterers.}\label{sparse signal}
\end{figure}

It is observed that in massive MIMO scenarios, the channel exhibits sparse multipath structure due to the limited scatterings. Thus, a virtual angular domain representation is applied to characterize the MIMO channel by fixed virtual receive and transmit directions \cite{43,53,57}:
\begin{equation}\label{ordinary model}
\normalsize
\mathbf{H}^{[i]}=\mathbf{U}\mathbf{\tilde{H}}^{[i]}\mathbf{V}^H,
\end{equation}
where $ \mathbf{\tilde{H}}^{[i]} \in \mathbb{C}^{N\times M} $ is the virtual angular domain representation of $\mathbf{H}^{[i]}$, $\mathbf{U}\in \mathbb{C}^{N\times N}$ and $\mathbf{V}\in \mathbb{C}^{M\times M}$ denote unitary spatial Fourier transform matrices for the angular domain transformation at the UE side and the BS side, respectively. The virtual angular representation channel matrix $\mathbf{\tilde{H}}^{[i]}$ describes the physical channel via fixed virtual angles with finite angular resolutions,  where the departure and arrival virtual angle grids are uniformly sampled with angular resolutions $M$ and $N$, respectively. The $(n,m)$-th element of $ \mathbf{\tilde{H}}^{[i]} $ is nonzero only if there exists a spatial path in which the signal departs from BS in the $m$-th angle grids and arrives at UE in the $n$-th angle grids. The value of the $(n,m)$-th element indicates the fading coefficient of the spatial path \cite{43}.  Besides, the channel matrix in angular domain has the following characteristics as indicated in \cite{9} and \cite{10}:

\begin{itemize}[leftmargin=1pt, labelsep=4pt,itemindent=10pt, listparindent=10pt]
\item \textbf{ Intra-frame sparsity for UE due to the limited scatterers at BS:} 
In the downlink massive MIMO system, there are limited scatterings at the BS side but rich local scatterings around UE when it is located at low elevation, thus all the multiple spatial paths are reflected by the rich local scatterings and arrive at UE in omni-directions  \cite{22}. Accordingly, each AoA of the channel matrix in the $i$-th frame corresponds to the same AoDs, i.e., $ supp(\mathbf{\tilde{H}}^{[i]}_{(1,:)})=\cdots= supp(\mathbf{\tilde{H}}^{[i]}_{(j,:)})=\cdots= supp(\mathbf{\tilde{H}}^{[i]}_{(N,:)}) \triangleq \mathbf{\Gamma}^{[i]}$, such sparse pattern is often termd as block sparsity \cite{10,56}. The example of $N=2$ is depicted for illustration in Fig. \ref{sparse signal}. As a frame-wise block sparse model, $|\mathbf{\Gamma}^{[i]}|$ is generally upper bounded as indicated by \cite{9,22}, where we assume that $|\mathbf{\Gamma}^{[i]}| \leq \overline{s}$. Furthermore, the nonzero elements in $\mathbf{\tilde{H}}^{[i]}$ are assumed to be i.i.d. complex Gaussian distributed with zero mean and unit variance \cite{9}.

\item \textbf{ Inter-frame sparsity among consecutive frames due to the shared scatterers at BS:} 
As depicted in Fig.\ref{system}, massive MIMO channels for consecutive frames are usually intercorrelated because of the shared scatterers at BS, which result in similar AoDs in consecutive frames \cite{9,22}. In order to accurately model such correlation, two kinds of inter-frame sparsity are explored. 
The first one is the small-scale inter-frame sparsity that models the correlation of AoDs that only appear in the channels between two adjacent frames, which can be represented by the shared support indices:
\begin{equation}\label{small inter common support set}
\overline{s} > | \mathbf{\Gamma}^{[i-1]} \cap \mathbf{\Gamma}^{[i]}| \geq s_c,
\end{equation}
where $s_c$ and $\overline{s}$ denote the upper bound and lower bound of such intersection set. The orange boxes depicted in Fig. \ref{sparse signal} illustrate this kind of correlation.

The second one is the large-scale inter-frame sparsity that models the correlation of AoDs that appear in the channels among $L$ consecutive frames received by UE, which can be represented by the common row support set:
\begin{equation}\label{inter common support set}
\mathbf{\Gamma}^s = \cap_{i=1}^{L} \mathbf{\Gamma}^{[i]},
\end{equation}
where its cardinality $|\mathbf{\Gamma}^s|=S$. Note that the indices in (\ref{inter common support set}) are overlapped with part of the indices in (\ref{small inter common support set}). That is, the small-scale inter-frame sparsity is a special case of large-scale inter-frame sparsity, where these two kinds of sparsity differ in terms of the number of the frames that are correlated. Therefore, for a clear demonstration of (\ref{inter common support set}), the blue boxes instead of orange boxes are depicted in Fig. \ref{sparse signal} to illustrate this kind of correlation with $L=3$ for illustration.
\end{itemize}

\subsection{Problem Formulation}
By applying virtual angular transformation, the model (\ref{CSmodel}) is rewritten as follows:
\begin{equation}\label{angular burst model}
\underbrace{(\mathbf{Y}^{[i]})^H \mathbf{U}}_{\mathbf{Z}^{[i]}}= \underbrace{ (\mathbf{X}^H \mathbf{V})}_{\mathbf{\Phi}} \underbrace{(\mathbf{\tilde{H}}^{[i]})^H}_{\mathbf{S}^{[i]}} + \underbrace{(\mathbf{N}^{[i]})^H \mathbf{U}}_{\mathbf{W}^{[i]}},
\end{equation}
where $\mathbf{Z}^{[i]}\in \mathbb{C}^{T\times N}$, $\mathbf{\Phi}\in \mathbb{C}^{T\times M}$, $\mathbf{W}^{[i]}\in \mathbb{C}^{T\times N}$ and $\mathbf{S}^{[i]}\in \mathbb{C}^{M\times N}$. Note that $\mathbf{S}^{[i]}$  is a block sparse matrix with only $|\mathbf{\Gamma}^{[i]}|$ nonzero rows, representing the intra-frame sparsity at BS side.

When a UE with $N$ antennas receives $L$ consecutive frames from BS, we concatenate the received signals as $\mathbf{R}=[\mathbf{Z}^{[1]},\cdots,\mathbf{Z}^{[L]}]$ in the form of:
\begin{equation}\label{overall angular burst model}
\mathbf{R}=\mathbf{\Phi} \mathbf{G} + \mathbf{N},
\end{equation}
where $\mathbf{G}=[\mathbf{S}^{[1]},\cdots,\mathbf{S}^{[L]}]$ and $\mathbf{N}=[\mathbf{W}^{[1]},\cdots,\mathbf{W}^{[L]}]$ denote the concatenated row-sparsity channel matrices and the noise across $L$ consecutive frames, respectively.

The optimization objective of massive MIMO channel estimation is modeled as follows:
\begin{equation}\label{overall objective}
\min_{\mathbf{G} \in \mathbb{C}^{M \times NL}}\frac{1}{2}||\mathbf{R}-\mathbf{\Phi}\mathbf{G}||_F^2 + \xi(\mathbf{G}),
\end{equation}
where $\xi(\mathbf{G})$ means the sparse regularizer on the concatenated row-sparse channel matrix $\mathbf{G}$.

However, it is hard to directly reconstruct $\mathbf{G}$ via the above optimization problem since no reasonable sparse regularizer $\xi(\cdot)$ can effectually capture the complicated sparse structure induced by the intra-frame sparsity and the two kinds of inter-frame sparsity simultaneously.  To solve the objective (\ref{overall objective}), a two-stage structure is proposed in this work which consists of the coarse estimation part and the fine correction part, where the large-scale and the small-scale inter-frame sparsities are exploited by these two parts, respectively.

Considering the large-scale inter-frame sparsity among consecutive frames, the corresponding channel coefficients form the nonzero rows in $\mathbf{G}$, i.e., the rows formed by blue boxes in Fig. \ref{sparse signal}. The regularizer $\xi(\cdot)$ is chosen to be $\xi(\mathbf{G})=||\mathbf{G}||_{2,1}$ to exploit the common row sparsity introduced by large-scale inter-frame correlation.  The objective of coarse estimation part is defined as follows: 
\begin{equation}\label{coarse objective}
\begin{aligned}
&\min_{\mathbf{G} \in \mathbb{C}^{M \times NL}}\frac{1}{2}||\mathbf{R}-\mathbf{\Phi}\mathbf{G}||_F^2 + \alpha ||\mathbf{G}||_{2,1},
\end{aligned}
\end{equation}
which is a standard $l_{2,1}$ minimization problem and $\alpha$ stands for the regularization parameter.

Moreover, based on Fig. \ref{sparse signal}, the real sparse channels are block sparse within one frame due to the intra-frame sparsity, and have shared support set between two adjacent frames due to the small-scale inter-frame sparsity. However, they cannot be fully captured by the mixed $l_{2,1}$ term in (\ref{coarse objective}), because the significance of those partially nonzero rows such as the yellow and orange boxes in Fig. \ref{sparse signal} are undetermined during the row-wise energy calculation and shrinkage of $\mathbf{G}$ through the mixed $l_{2,1}$ penalty, which definitely affects the estimation performance. Thus, we turn to design a fine correction part to combine both sparsities to refine the estimated results in (\ref{coarse objective}).

To capture the intra-frame sparsity and small-scale inter-frame sparsity, we aim to refine the coarsely estimated channel in a frame-wise way and exploit the small-scale inter-frame sparsity in the form of partial support information inspired by \cite{9,24,54,55}. The following weighted $\ell_{2,1}$ minimization problem of the fine correction part is defined as follows:
\begin{equation}\label{weighted fine objective}
\min_{\mathbf{S}^{[i]}\in \mathbb{C}^{M\times N}}\frac{1}{2}||\mathbf{Z}^{[i]}-\mathbf{\Phi}\mathbf{S}^{[i]}||_F^2 + \lambda \sum_{j=1}^M \omega_j ||\mathbf{S}_{(j,:)}^{[i]}||_{2}
\end{equation}
with
\begin{equation}\nonumber
\begin{aligned}
&\ \omega_{j}=
\begin{cases}
1, & j \notin \hat{\mathbf{\Gamma}}^{[i-1]},\\
\omega, & j \in \hat{\mathbf{\Gamma}}^{[i-1]},
\end{cases}
\end{aligned}
\end{equation}
The parameter $\lambda$ stands for the regularization parameter. $\omega$ is a training parameter learned from massive data in our paper. The value of $\omega$ determines the degree of correlation between the channels in the $i$-th frame and the $i-1$-th frame. A smaller $\omega$ corresponds to a higher probability that the $j$-th row of $\mathbf{S}^{[i]}$ is also nonzero row as in $\hat{\mathbf{\Gamma}}^{[i-1]}$. The set $\hat{\mathbf{\Gamma}}^{[i-1]}$ is known as priori support information to help reconstruct the channel in the $i$-th frame, where $\hat{\mathbf{\Gamma}}^{[i-1]}$ is usually acquired through the reconstruction of $\mathbf{S}^{[i-1]}$.

\section{PROPOSED SCHEMES}
\begin{figure*}
	\centering
   \includegraphics[scale=0.43]{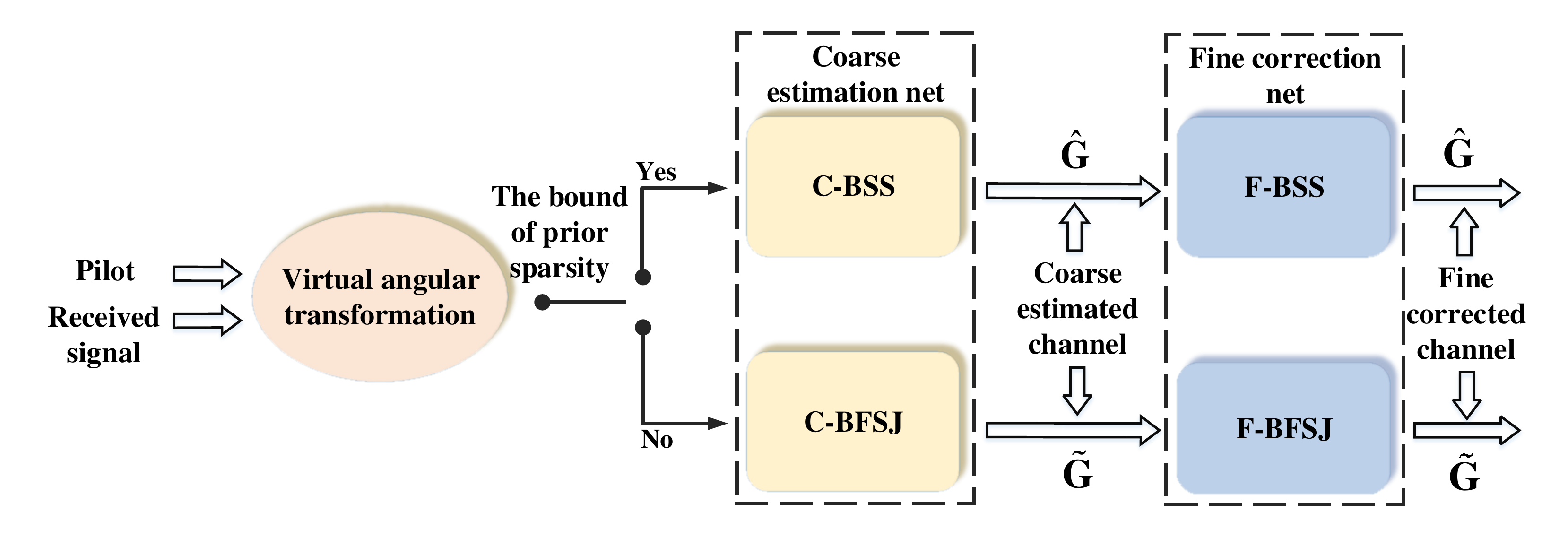}
	\caption{The whole network structure of the proposed schemes.}
	\label{whole network}
\end{figure*}

In this section, we describe the two proposed schemes that solve both the objective (\ref{coarse objective}) and (\ref{weighted fine objective}). The two-stage structure of the proposed schemes is depicted in Fig. \ref{whole network}, which includes coarse estimation net and fine correction net. The requirement of priori sparsity bound information is the only difference between these two schemes. Moreover, in order to better illustrate them, we first introduce a thresholding function that requires priori sparsity bound information, then this function is extended to eliminate such requirement. Next, the basic sub-networks of our schemes, i.e., coarse estimation net and fine correction net, are respectively realized to solve (\ref{coarse objective}) and (\ref{weighted fine objective}) based on the proposed thresholding functions.

\begin{figure}[t]
	\centering
 \subfigure[Coarse estimation net.]{
    \begin{minipage}[t]{1\linewidth}
 \begin{center}
    \includegraphics[scale=0.62]{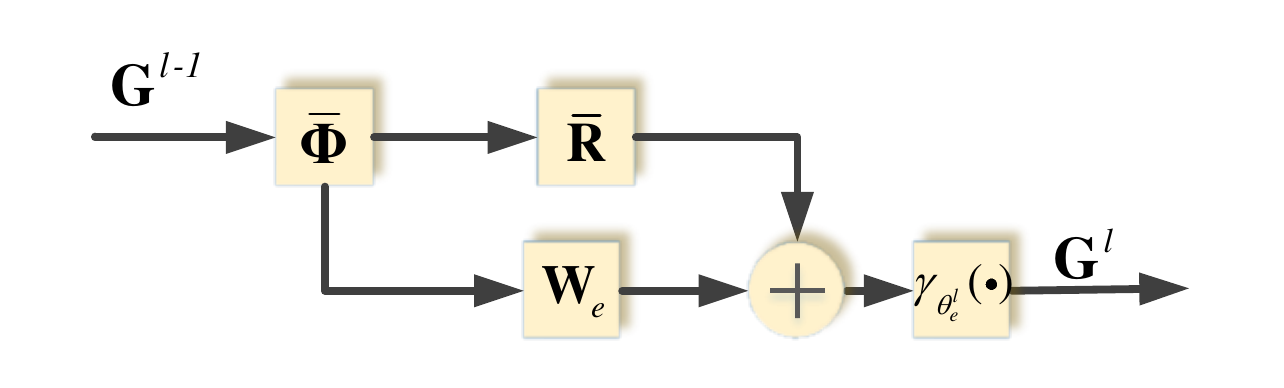}
 \label{coarse net}
 \end{center}
 \end{minipage}}
 \subfigure[Fine correction net.]{
    \begin{minipage}[t]{1\linewidth}
 \begin{center}
    \includegraphics[scale=0.62]{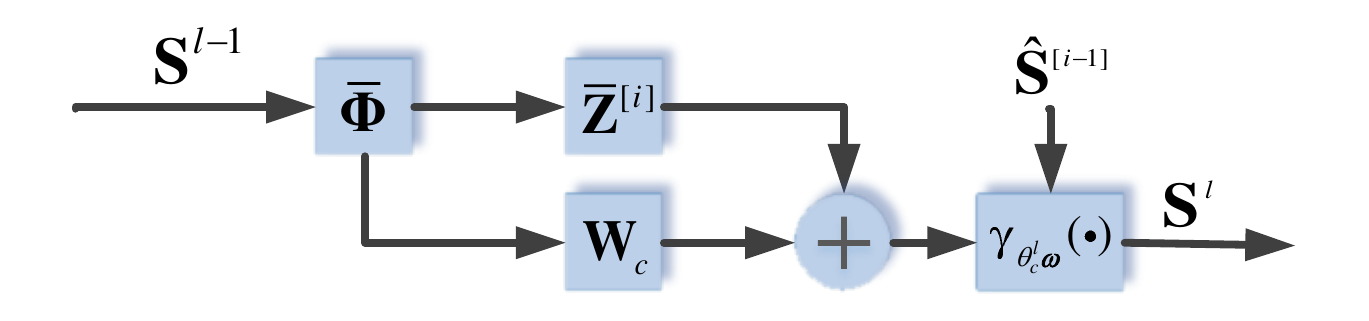}
 \label{fine net}
 \end{center}
 \end{minipage}}
 \caption{Illustration of the network structure. (a) The structure of each layer in coarse estimation net. (b) The structure of each layer in fine correction net.}
\label{detailed network}
\end{figure}
\subsection{Proposed Thresholding Functions}
The ISTA-based algorithms in \cite{10,21,26} are widely applied to solve $l_{2,1}$ minimization problem.
However, its estimated sparse coefficients are not accurate because of the attenuation caused by its shrinkage factor \cite{51}. Therefore, in this section, two novel thresholding functions are proposed to solve this issue.

To begin with, we introduce block coordinate descent with multiple measurement vector (BCD-MMV) in \cite{21}, which originates from ISTA to solve $l_{2,1}$ minimization problem, i.e., (\ref{weighted fine objective}) when $\omega=1$ and (\ref{coarse objective}). The $l$-th iteration of BCD-MMV to solve (\ref{weighted fine objective}) when $\omega=1$ is presented as follows:
\begin{equation}\label{BCD_initial}
\mathbf{S}^l  = \delta_{\lambda/q}(\mathbf{S}^{l-1} + \mathbf{\Phi}^T(\mathbf{Z}^{[i]}-\mathbf{\Phi}\mathbf{S}^{l-1})).\\
\end{equation}
In the above equation, $\mathbf{S}^l$ is the estimation of the ground truth $\mathbf{S}^{[i]}$, $q$ is usually chosen as the largest eigenvalue of $\mathbf{\Phi}^T \mathbf{\Phi}$ and the function $\delta_{\lambda/q}(\cdot)$ is defined as:
\begin{equation}\label{BCD thresholding}
\delta_{\lambda/q}(\mathbf{V}_{(j,:)}) = 
\left\{
\begin{aligned}
&\frac{\mathbf{V}_{(j,:)}(||\mathbf{V}_{(j,:)}||_2 - \lambda/q)}{||\mathbf{V}_{(j,:)}||_2}, &  ||\mathbf{V}_{(j,:)}||_2 > \lambda/q,\\
&\mathbf{0}_{1\times N}, &  ||\mathbf{V}_{(j,:)}||_2 < \lambda/q.\\
\end{aligned}
\right.\\
\end{equation}

This function (\ref{BCD thresholding}) adaptively chooses nonzero entries by comparing their row norms with the predefined parameter $\lambda/q$. However, according to \cite{59}, the magnitude of estimated channel coefficients are attenuated by the shrinkage factor in (\ref{BCD thresholding}):
\begin{equation}\label{shrink factor}
\frac{(||\mathbf{V}_{(j,:)}||_2 - \lambda/q)}{||\mathbf{V}_{(j,:)}||_2},
\end{equation}
which is caused by the presence of regularization term in $l_{2,1}$ minimization problem \cite{51}, such as  $\sum_{j=1}^M ||\mathbf{S}_{(j,:)}^{[i]}||_{2}$ in (\ref{weighted fine objective}) when $\omega=1$.

Inspired by \cite{19}, we develop a block thresholding function with support selection (BSS) to eliminate such negative effect, where we generalize the support selection (SS) thresholding function in \cite{19} to solve the $l_{2,1}$ minimization problem:
\begin{equation}\label{soft thre support selection function}
\gamma_{\mathbf{\theta}^l}^{ss}(\mathbf{V}_{(j,:)}, \mathbf{\Omega}_{l}^{ss}) = \frac{\mathbf{V}_{(j,:)}}{||\mathbf{V}_{(j,:)}||_2} \eta^{ss}_{\mathbf{\theta}^l }(\mathbf{V}_{(j,:)}, \mathbf{\Omega}_{l}^{ss}),
\end{equation}
where
\begin{equation}\label{right part of soft thre support selection function}
\begin{aligned}
&\eta_{\mathbf{\theta}^l}^{ss}(\mathbf{V}_{(j,:)}, \mathbf{\Omega}_{l}^{ss}) = \\
&\left\{
\begin{aligned}
&||\mathbf{V}_{(j,:)}||_2, & {\rm if}\ & ||\mathbf{V}_{(j,:)}||_2 > \mathbf{\theta}^l, j \in \mathbf{\Omega}_{l}^{ss}, \\
&||\mathbf{V}_{(j,:)}||_2 - \mathbf{\theta}^l ,& {\rm if}\ &  ||\mathbf{V}_{(j,:)}||_2 > \mathbf{\theta}^l, j \notin \mathbf{\Omega}_{l}^{ss}, \\
&0,& {\rm if}\ & ||\mathbf{V}_{(j,:)}||_2 \leq \mathbf{\theta}^l.\\
\end{aligned}
\right.\\
\end{aligned}
\end{equation}
and $\theta^1 = \cdots = \theta^L =\lambda/q$. The set $\mathbf{\Omega}_{l}^{ss}$ includes the elements with the largest $p_{_l}\%$ magnitudes in $\lbrace ||\mathbf{V}_{(j,:)}||_2\ |\  j=1,\cdots,M\rbrace$:
\begin{equation}\label{support selection set}
\mathbf{\Omega}_{l}^{ss} = \lbrace j_i,\cdots,j_{\lfloor p_lM \rfloor}\ | \ ||\mathbf{V}_{(j_1,:)}||_2 \geq \cdots \geq ||\mathbf{V}_{(j_M,:)}||_2 \rbrace.
\end{equation} 
The following equation is proposed to calculate the value of $p_{_l}$:
\begin{equation}\label{ss thresholding}
p_l = (\frac{p_{min}}{M} + \frac{p_{max}-p_{min}}{M(L-1)} (l-1))\times100\%, l=1,\cdots,L.
\end{equation}
where $L$ stands for the number of layers in the corresponding network. The above equation indicates that $p_l \in [p_{min}/M, p_{max}/M]$, where $p_{max}$ and $p_{min}$ denote the upper bound and the lower bound of the cardinality of $\mathbf{S}^{[i]}$, respectively.

According to the function (\ref{BCD thresholding}), the larger value of $||\mathbf{V}_{(j,:)}||_2$, the higher probability that $j$-th row $\mathbf{V}_{(j,:)}$ is reconstructed as a nonzero row. Therefore, the entries included in $\mathbf{\Omega}_{l}^{ss}$ are those with the highest probability to be nonzeros. In the proposed function (\ref{right part of soft thre support selection function}), we trust the entries in $\mathbf{\Omega}_{l}^{ss}$ as `true support entries' and remove the magnitude attenuation corresponding to those entries in the first term of (\ref{right part of soft thre support selection function}), by reconstructing their coefficients as $\mathbf{V}_{(j,:)}$ instead of $\frac{\mathbf{V}_{(j,:)}(||\mathbf{V}_{(j,:)}||_2 - \lambda/q)}{||\mathbf{V}_{(j,:)}||_2}$. Thus, the proposed BSS function demonstrates advantages on channel reconstruction accuracy.

However, such thresholding function may not be available in practical scenario since the choice of $p_l$ in (\ref{ss thresholding}) requires the priori sparsity bound information $p_{max}$ and $p_{min}$. For this issue, we exploit the technique called `first significant jump' (FSJ)\cite{26} to eliminate such requirement.

The proposed block thresholding function with `first significant jump' (BFSJ) is defined as follows:
\begin{equation}\label{soft thre first signifdicant jump function}
\gamma_{\mathbf{\theta}^l}^{fir}(\mathbf{V}_{(j,:)},\mathbf{\Omega}_{l}^{fir}) = \frac{\mathbf{V}_{(j,:)}}{||\mathbf{V}_{(j,:)}||_2} \eta^{fir}_{\mathbf{\theta}^l }(\mathbf{V}_{(j,:)}, \mathbf{\Omega}_{l}^{fir}).
\end{equation}

The generation procedures of $\mathbf{\Omega}_{l}^{fir}$ are described as follows:
\begin{enumerate}[label={(\arabic*)}]
    \item Sort the norm $||\mathbf{V}_{(j,:)}||_2$  in increasing order, i.e., $||\mathbf{V}_{(j_1,:)}||_2<\cdots<||\mathbf{V}_{(j_k,:)}||_2<\cdots<||\mathbf{V}_{(j_M,:)}||_2$, where $j_k$ stands for the row index whose corresponding $||\mathbf{V}_{(j_k,:)}||_2$ is the $k$-th largest.
    
    \item Find the smallest $k$ that satisfies:
    \begin{equation}\nonumber
    ||\mathbf{V}_{(j_{k+1},:)}||_2 - ||\mathbf{V}_{(j_k,:)}||_2 > \frac{||\mathbf{V}_{(j_M,:)}||_2}{T},
    \end{equation}
    then we set $\beta = ||\mathbf{V}_{(j_k,:)}||_2$.

    \item Finally, we set $\mathbf{\Omega}_{l}^{fir} =\lbrace j: ||\mathbf{V}_{(j,:)}||_2 > \beta \rbrace $.
\end{enumerate}

Note that we choose $\mathbf{\Omega}_{l}^{fir}$ based on the thresholding parameter $\beta=||\mathbf{V}_{(j_k,:)}||_2$ instead of $p_l$ in (\ref{soft thre support selection function}), thus the requirement of priori information is eliminated. The parameter $\beta$ can properly decide the set $\mathbf{\Omega}_{l}^{fir}$ corresponding to the selected nonzero entries based on the following reasons: the true nonzero entries $\lbrace ||\mathbf{V}_{(j,:)}||_2 \ |\  j \in \mathbf{\Gamma}^{[i]} \rbrace$ have magnitudes larger than zero while the true zero entries $\lbrace ||\mathbf{V}_{(j,:)}||_2 \ |\  j \in (\mathbf{\Gamma}^{[i]})^c \rbrace$ have magnitudes that approach zero. After sorting $\lbrace ||\mathbf{V}_{(j,:)}||_2 \ |\  j \in \lbrace 1,2,\cdots,M\rbrace \rbrace$ in an increasing order, we note that there exists a sudden change point of magnitude in the sorted sequence. The index of this point, denoted as $j_k$, will be chosen as the boundary index. Meanwhile, the corresponding $l_2$ norm $||\mathbf{V}_{(j_{k},:)}||_2$ is set as the boundary value $\beta$, where we consider the entries with norm $\lbrace ||\mathbf{V}_{(j,:)}||_2 \ |\  j \in \lbrace 1,2,\cdots,M\rbrace \rbrace$ larger than $\beta$ as the true nonzero entries, and the entries with norm smaller than $\beta$ as true zero elements \cite{26}.

In a word, the proposed BFSJ function replaces $\mathbf{\Omega}_{l}^{ss}$ with the set $\mathbf{\Omega}_{l}^{fir}$, which enjoys the advantage of no requirement of prior sparsity bound information without drastically degrading the performance.

\subsection{Coarse Estimation Net}\label{coarse estimation net}
In this section,  we propose coarse estimation net to coarsely estimate the channel coefficients with the optimization objective in (\ref{coarse objective}), where large-scale inter-frame sparsity is exploited. To the best of our knowledge, we are the first to combine such sparsity with the deep unrolling technique to improve the channel estimation performance.  Firstly, the coarse estimation net with the requirement of prior sparsity bound information is presented for illustration, i.e., coarse estimation net with BSS (C-BSS), then we propose the network that requires no prior sparsity bound information, i.e., coarse estimation net with BFSJ (C-BFSJ). The detailed network structure of coarse estimation net is depicted in Fig. \ref{coarse net}.

We first give the real-valued counterpart of (\ref{overall angular burst model}) as:
\begin{equation}\label{coarse angular real model}
\begin{aligned}
\bar{\mathbf{R}}&=\bar{\mathbf{\Phi}} \bar{\mathbf{G}}+ \bar{\mathbf{N}}\\
&=\begin{bmatrix} \mathbbm{R}(\mathbf{\Phi}) & -\mathbbm{I}(\mathbf{\Phi})  \\ \mathbbm{I}(\mathbf{\Phi}) & \mathbbm{R}(\mathbf{\Phi}) \end{bmatrix}    \begin{bmatrix} \mathbbm{R}(\mathbf{G})  \\ \mathbbm{I}(\mathbf{G}) \end{bmatrix} +  \begin{bmatrix} \mathbbm{R}(\mathbf{N})  \\ \mathbbm{I}(\mathbf{N}) \end{bmatrix},
\end{aligned}
\end{equation}
where $\bar{\mathbf{R}}$, $\bar{\mathbf{\Phi}}$, $\bar{\mathbf{G}}$ and $\bar{\mathbf{N}}$ denote the real-valued version of received signal, measurement matrix, actual channel matrix and noise, respectively.

The $l$-th iteration of BCD-MMV algorithm to solve (\ref{coarse objective}) is:
\begin{equation}\label{BCD}
\mathbf{G}^l  = \delta_{\alpha/q}(\mathbf{G}^{l-1} + \bar{\mathbf{\Phi}}^T(\bar{\mathbf{R}}-\bar{\mathbf{\Phi}}\mathbf{G}^{l-1})),\\
\end{equation}
where $\mathbf{G}^l$ is the estimation of the ground truth $\bar{\mathbf{G}}$ in the $l$-th iteration.  The unrolled structure of the coarse estimation net in \textbf{Network} \ref{alg:coarse estimation net} is realized  based on (\ref{BCD}), where $\bar{\mathbf{\Phi}}^T$ and $\alpha/q$ in (\ref{BCD}) are replaced by the trainable weights $\mathbf{W}^l_{e}$ and $\theta_e^l$ in Step 3 of \textbf{Network} \ref{alg:coarse estimation net}, respectively.  Note that by considering large-scale inter-frame sparsity, (\ref{coarse objective}) is a typical $l_{2,1}$ minimization problem,  thus $\delta_{\alpha/q}(\cdot)$ can be directly extended to (\ref{soft thre support selection function}) and (\ref{soft thre first signifdicant jump function}).

\begin{algorithm}[t]
	\renewcommand{\algorithmicrequire}{\textbf{Input:}}
	\renewcommand{\algorithmicensure}{\textbf{Output:}}
	\caption{Coarse Estimation Net}
	\label{alg:coarse estimation net}
	\begin{algorithmic}[1]
		\REQUIRE $\bar{\mathbf{R}}$, $\bar{\mathbf{\Phi}}$, $S$, \rm the number of layers$\ L_{e}$
		\ENSURE $\tilde{\mathbf{G}}$
        \STATE $\mathbf{Initialization:}$ $l=1$, $\mathbf{G}^0=\mathbf{0}_{M\times LN}$
        \WHILE {$l\leq L_{e}$ }
		\STATE $\mathbf{G}^l=\gamma_{\theta_e^l}(\mathbf{G}^{l-1} + \mathbf{W}^l_{e}(\bar{\mathbf{R}}-\bar{\mathbf{\Phi}}\mathbf{G}^{l-1}),\mathbf{\Omega}_l)$
		\STATE $l=l+1$
        \ENDWHILE
		\STATE \textbf{return}  $\tilde{\mathbf{G}}=\mathbf{G}^{L_{e}}=[\tilde{\mathbf{S}}^{[1]},\cdots,\tilde{\mathbf{S}}^{[L]}]$
	\end{algorithmic}
\end{algorithm}

In the case of MIMO channel estimation with known sparsity bound information, thresholding function (\ref{soft thre support selection function}) is utilized in \textbf{Network} \ref{alg:coarse estimation net}, where the corresponding net is termed as C-BSS. The iterative structure of C-BSS is defined as follows:

\begin{equation}\label{C-BSS}
\mathbf{G}^l  = \gamma_{\theta_e^l}^{ss}(\mathbf{G}^{l-1} + \mathbf{W}^l_{e}(\bar{\mathbf{R}}-\bar{\mathbf{\Phi}}\mathbf{G}^{l-1}), \mathbf{\Omega}^{ss}_l).
\end{equation}
Besides, for the setting of $p_l$ in (\ref{ss thresholding}) to control $\mathbf{\Omega}^{ss}_l$, the values of $p_{min}$, $p_{max}$ and $L$ are set as $s_c$, $S$ and $L_e$, respectively.

Unfortunately, the precise value of $S$ is unknown in practice. Therefore, with the given sparsity bound information $(\overline{s},s_c)$, we set the value of $S$ as the average row sparsity in $\bar{\mathbf{G}}$ instead:

\begin{equation}\label{coarse S}
S = 2b^{L-1}\mathbbm{E}(|\mathbf{\Gamma}^{[1]}|) + \frac{2a(1-b^{L-1})}{1-b},
\end{equation}
where
\begin{equation}\nonumber
\begin{aligned}
& a = \frac{(\mathbbm{E}(|\mathbf{\Gamma}^{[1]}|)-s_c)M}{M-|\mathbf{\Gamma}^{[1]}|},
& b = \frac{M-2\mathbbm{E}(|\mathbf{\Gamma}^{[1]}|)+s_c}{M-\mathbbm{E}(|\mathbf{\Gamma}^{[1]}|)}.
\end{aligned}
\end{equation}
Define the rows that have nonzero elements are nonzero rows, then the above equation indicates the average number of nonzero rows in $\bar{\mathbf{G}}$ by assuming that $\mathbbm{E}(|\mathbf{\Gamma}^{[1]}|) = \cdots = \mathbbm{E}(|\mathbf{\Gamma}^{[L]}|)=\frac{\bar{s}+s_c}{2}$. The detailed calculation is given in Appendix A.

Considering that C-BSS net requires the prior information of sparsity $(S,s_c)$, we further propose C-BFSJ net to eliminate such requirement. Note that the whole network structure of C-BFSJ are the same as C-BSS, thus the detailed explanation of C-BFSJ is omitted here for simplicity. C-BFSJ differs from C-BSS only in the utilization of the thresholding function (\ref{soft thre first signifdicant jump function}).

Based on the proposed coarse estimation net, we conclude that the way of estimating channel coefficients corresponding to large-scale inter-frame sparsity: finding the common support set $\mathbf{\Gamma}^s$ and estimating the corresponding coefficients. Firstly, we expect that $\lbrace ||\mathbf{V}^{est}_{(j,:)}||_2 \ |\ j \in \mathbf{\Gamma}^s \rbrace$ would be larger than $\lbrace ||\mathbf{V}^{est}_{(j,:)}||_2 \ |\ j \notin \mathbf{\Gamma}^s \rbrace$ with higher probability. Therefore, we choose the trainable threshold $\theta_e^l$ as boundary to indicate whether the $j$-th row $\mathbf{V}^{est}_{(j,:)}$ belongs to $\mathbf{\Gamma}^s$ or not. The rows whose $l_2$ norm are larger than $\theta_e^l$ are considered to constitute the following estimated common support set:
\begin{equation}
\tilde{\mathbf{\Gamma}}^s = \lbrace j \ | \ ||\mathbf{V}^{est}_{(j,:)}||_2 \geq \theta_e^l \rbrace,
\end{equation}
where 
\begin{equation}\label{coarse estimation gradient}
\mathbf{V}^{est} = \mathbf{G}^{l-1} + \mathbf{W}^l_{e}(\bar{\mathbf{R}}-\bar{\mathbf{\Phi}}\mathbf{G}^{l-1}).
\end{equation}
Secondly, the coefficients of these selected nonzero rows corresponding to $\tilde{\mathbf{\Gamma}}^s$ are derived based on (\ref{soft thre support selection function}) or (\ref{soft thre first signifdicant jump function}).

The above coarse estimation net derives the channel estimates, but the true channels often do not match with such estimated channels. The reason behind this is that the nonzero coefficients corresponding to small-scale inter-frame sparsity and intra-frame sparsity are undetermined during the shrinkage of $\mathbf{G}^l$ in (\ref{soft thre support selection function}) and (\ref{soft thre first signifdicant jump function}). The simple $l_{2,1}$ penalty is incapable of describing such complex pattern in $\mathbf{G}$.  For example, the channel coefficients corresponding to the yellow and orange boxes in Fig. \ref{sparse signal} can not be effectively reconstructed and are often misjudged as zeros in \textbf{Network} \ref{alg:coarse estimation net}. 

\subsection{Fine Correction Net}\label{fine correction net}
In this section, fine correction net is proposed to refine the coarsely estimated channel frame by frame with the optimization objective (\ref{weighted fine objective}), by utilizing the generalized version of BSS and BFSJ functions which exploit small-scale inter-frame sparsity between adjacent frames. To the best of our knowledge, we are the first to combine small-scale inter-frame sparsity with deep unrolling techniques, and also the first to combine coarse estimation net and fine correction net to greatly reduce the required pilot overhead. Besides, two versions of fine correction net are proposed. Firstly, the version that requires priori sparsity bound information is proposed based on the generalized BSS function, i.e., fine correction net with BSS (F-BSS). Then the version that eliminates such requirement is presented, i.e., fine correction net with BFSJ (F-BFSJ).  The detailed network structure of fine correction net is depicted in Fig. \ref{fine net}.

We first give the real-valued counterpart of (\ref{angular burst model}) in \textbf{Network} \ref{alg:fine correction net}: 
\begin{equation}\label{fine angular real model}
\begin{aligned}
\bar{\mathbf{Z}}^{[i]}&=\bar{\mathbf{\Phi}} \bar{\mathbf{S}}^{[i]}+\bar{\mathbf{W}}^{[i]}\\
&=\begin{bmatrix} \mathbbm{R}(\mathbf{\Phi}) & -\mathbbm{I}(\mathbf{\Phi})  \\ \mathbbm{I}(\mathbf{\Phi}) & \mathbbm{R}(\mathbf{\Phi}) \end{bmatrix}    \begin{bmatrix} \mathbbm{R}(\mathbf{S}^{[i]})  \\ \mathbbm{I}(\mathbf{S}^{[i]}) \end{bmatrix} +  \begin{bmatrix} \mathbbm{R}(\mathbf{W}^{[i]})  \\ \mathbbm{I}(\mathbf{W}^{[i]}) \end{bmatrix},
\end{aligned}
\end{equation}
where $\bar{\mathbf{\Phi}}$, $\bar{\mathbf{Z}}^{[i]}$ and $\bar{\mathbf{S}}^{[i]}$ denote the real-valued version of measurement matrix, the pilot signal and the channel in the $i$-th frame, respectively. 
The unrolled structure of fine correction net in \textbf{Network} \ref{alg:fine correction net} is realized based on (\ref{BCD_initial}), where the trainable weight $\omega$ is added into this net to incorporate small-sacle inter frame sparsity. Meanwhile, $\bar{\mathbf{\Phi}}^{T}$ and $\lambda/q$ in (\ref{BCD_initial}) are replaced by the trainable weights $\mathbf{W}^l_{c}$ and $\mathbf{\theta}_c^l$ in Step 4 of \textbf{Network} \ref{alg:fine correction net}, respectively \cite{19}.
\begin{algorithm} [t]
	\renewcommand{\algorithmicrequire}{\textbf{Input:}}
	\renewcommand{\algorithmicensure}{\textbf{Output:}}
	\caption{Fine Correction Net}
	\label{alg:fine correction net}
	\begin{algorithmic}[1]
		\REQUIRE  \rm The number of layers$\ L_{c}$, \rm the number of frames $L$, $\bar{\mathbf{\Phi}}$, $\lbrace \bar{\mathbf{Z}}^{[i]}| i=1,\cdots,L \rbrace$ and coarsely estimated channels $\lbrace \tilde{\mathbf{S}}^{[i]}| i=1,\cdots,L \rbrace$.
		\ENSURE $\hat{\mathbf{G}}$
		\WHILE {$i\leq L$ }
        \STATE $\mathbf{Initialization:}$ $l=1$, $\mathbf{S}^0=\tilde{\mathbf{S}}^{[i]}$, $\hat{\mathbf{\Gamma}}^{[0]}=\emptyset$ and $\hat{\mathbf{\Gamma}}^{[i-1]}={supp(\hat{\mathbf{S}}^{[i-1]})}$\\
        \WHILE {$l\leq  L_{c}$}
		\STATE $\mathbf{S}^l= \gamma_{\mathbf{\theta}_c^l \bm{\omega}} (\mathbf{S}^{l-1} + \mathbf{W}^l_{c}(\bar{\mathbf{Z}}^{[i]}-\bar{\mathbf{\Phi}}\mathbf{S}^{l-1}),\mathbf{\Omega}_l)$
		\STATE $\bm{\omega}=[\omega_1,\cdots,\omega_M]$, where $\omega_j = 
				\left\{
				\begin{aligned}
				&\omega : & \ j \in \hat{\mathbf{\Gamma}}^{[i-1]} \\
				& 1 : & \  j \notin \hat{\mathbf{\Gamma}}^{[i-1]}\\
				\end{aligned}
				\right.$
		\STATE $l=l+1$
        \ENDWHILE
		\STATE $\hat{\mathbf{S}}^{[i]} = \mathbf{S}^{ L_{c}}$
		\ENDWHILE
		\STATE \textbf{return}  $\hat{\mathbf{G}}=[\hat{\mathbf{S}}^{[1]},\cdots,\hat{\mathbf{S}}^{[L]}] $
	\end{algorithmic}
\end{algorithm}

Firstly, we introduce F-BSS net which requires the priori sparsity bound information. In this net, BSS thresholding function to optimize $l_{1,2}$ minimization problem is generalized to solve weighted $l_{2,1}$ minimization problem (\ref{weighted fine objective}), where small-scale inter-frame correlation is exploited as weight $\omega$. The generalized BSS function is defined as follows:
\begin{equation}\label{soft weighted thre support selection function}
\gamma_{\mathbf{\theta}_c^l  \omega_j}^{ss}(\mathbf{V}^{cor}_{(j,:)},\mathbf{\Omega}_{l}^{ss}) = \frac{\mathbf{V}^{cor}_{(j,:)}}{||\mathbf{V}^{cor}_{(j,:)}||_2} \eta_{\mathbf{\theta}_c^l  \omega_j}^{ss}(\mathbf{V}^{cor}_{(j,:)},\mathbf{\Omega}_{l}^{ss})
\end{equation}
\begin{equation}\nonumber
\begin{aligned}
&\eta_{\mathbf{\theta}_c^l \omega_j}^{ss}(\mathbf{V}^{cor}_{(j,:)},\mathbf{\Omega}_{l}^{ss}) = \\
&\left\{
\begin{aligned}
&||\mathbf{V}^{cor}_{(j,:)}||_2, & {\rm if}\ & ||\mathbf{V}^{cor}_{(j,:)}||_2 > \mathbf{\theta}_c^l, j \in \mathbf{\Omega}_{l}^{ss}, \\
&||\mathbf{V}^{cor}_{(j,:)}||_2 - \mathbf{\theta}_c^l \omega_j,& {\rm if}\ &  ||\mathbf{V}^{cor}_{(j,:)}||_2 > \mathbf{\theta}_c^l \omega_j, j \notin \mathbf{\Omega}_{l}^{ss}, \\
&0,& {\rm if}\ & ||\mathbf{V}^{cor}_{(j,:)}||_2 \leq \mathbf{\theta}_c^l\omega_j.\\
\end{aligned}
\right.\\
\end{aligned}
\end{equation}
where
\begin{equation}\label{fine matrix multiplication}
\mathbf{V}^{cor} = \mathbf{S}^{l-1} + \mathbf{W}^l_{c}(\bar{\mathbf{Z}}^{[i]}-\bar{\mathbf{\Phi}}\mathbf{S}^{l-1}).
\end{equation}
For the setting of $p_{_l}$ in (\ref{ss thresholding}) to control $\mathbf{\Omega}_{l}^{ss}$, we set $p_{min}$, $p_{max}$ and $L$ as $s_c$, $\bar{s}$ and $L_c$, respectively. $\omega$ is a trainable parameter that indicates the degree of small-scale inter-frame correlation between adjacent frames.

In \textbf{Network} \ref{alg:fine correction net}, $\hat{\mathbf{\Gamma}}^{[i-1]}=supp(\hat{\mathbf{S}}^{[i-1]})$ is utilized as partial support information after the reconstruction of $\hat{\mathbf{S}}^{[i-1]}$. The parameter $\theta_c^l$ is multiplied by weight $\omega$ which belongs to $\hat{\mathbf{\Gamma}}^{[i-1]}$ to decrease its vaule in (\ref{soft weighted thre support selection function}), leading to a higher probability that the $\hat{\mathbf{S}}^{[i]}_{(j,:)}$ is reconstructed as nonzero row.

According to the above explanation, the initial value for the trainable weight $\omega$ is essential and needs to be properly selected. The value of $\omega$ is set to be inversely proportional to two factors: The first is the small-scale inter-frame correlation between $\mathbf{\Gamma}^{[i-1]}$ and $\mathbf{\Gamma}^{[i]}$, and the second is the reconstruction quality of $\hat{\mathbf{\Gamma}}^{[i-1]}$. The higher correlation or the higher reconstruction quality of the sparse MIMO channels, the smaller initial value of $\omega$ we should set and the more accurate sparse channels can be estimated. Additionally, the range of its initial value is within $[0,1]$ as mentioned in (\ref{weighted fine objective}).

Secondly, F-BFSJ network is proposed to eliminate the requirement of sparsity bound information $(\overline{s},s_c)$ in F-BSS network. The whole network structure of F-BFSJ is the same as F-BSS except that we replace the set $\mathbf{\Omega}_{l}^{ss}$ in (\ref{soft weighted thre support selection function}) with the set $\mathbf{\Omega}_{l}^{fir}$ generated by `first significant jump' rule. 

In a word, the coarse estimation net and the fine correction net differ from each other in the ways of exploiting different inter-frame sparsities. The coarse estimation net exploits the large-scale inter-frame sparsity across $L$ consecutive frames, but ignores the channel coefficients affected by the intra-frame sparsity and the small-scale inter-frame sparsity. Meanwhile, the fine correction net refines the coarse channel estimations frame by frame with the small-scale inter-frame sparsity between adjacent frames, where large-scale effect is ignored as the fine correction net only operates between two adjacent received frames. Only the combination of the two nets can exploit both large-scale and small-scale inter-frame sparsity of massive MIMO channels. Further simulation results in Section ${\rm\uppercase\expandafter{\romannumeral4}}$ 
are provided to verify that such concatenation is necessary for guaranteeing performance.

\section{SIMULATION RESULTS}
We consider a downlink massive MIMO system with one BS and one UE, where BS and UE are equipped with $M=128$ and $N=2$ antennas, respectively. The angular domain channel coefficients in $\mathbf{G}$ are i.i.d. complex Gaussian with zero mean and unit variance as those in \cite{9}. Suppose that the number of spatial paths from BS to UE is randomly generated as $|\mathbf{\Gamma}^{[i]}| \sim \mathcal{Z}(\overline{s}-3,\overline{s}-1)$, the number of shared paths between consecutive frames obeys $| \mathbf{\Gamma}^{[i-1]} \cap \mathbf{\Gamma}^{[i]}| \sim \mathcal{Z}(s_c,s_c+1)$, where $\mathcal{Z}(a,b)$ denotes integer uniform distribution between integer $a$ and integer $b$. The number of frames that are jointly considered is $L=7$. The angle of departure is uniformly distributed over $(0, \pi)$, indicating the equal probability of each column in $\mathbf{\tilde{H}}^{[i]}$ to be nonzero column. The elements in the pilot matrix are i.i.d. and drawn from the uniform distribution $\mathcal{U}(-\sqrt{1/M},\sqrt{1/M})$.

Note that the coarse estimation net and the fine correction net are trained layer by layer as in \cite{19}. At the training stage, the parameters  $\lbrace \mathbf{W}^l_{e}, \theta_e^l \rbrace ^{L_{e}}_{l=1}$ in the coarse estimation net are trained first but are frozen when we train the parameters $\lbrace \mathbf{W}^l_{c}, \theta_c^l, \omega \rbrace ^{L_{c}}_{l=1}$ in the fine correction net. Besides, we find that after respective training of coarse and fine nets, fine-tune step with both the two nets brings no obvious performance improvements. Thus, we omit this fine-tune step of the proposed schemes for simplicity.  The number of layers for coarse estimation net $L_{e}$ and fine correction net $L_{c}$ are set as $8$ and $16$, respectively.  Therefore, the number of layers of all the deep unrolling network listed below (including the variants of the proposed structure and other existing deep unrolling networks) is also $24$ for a fair comparison. The initial input of our proposed schemes and existing schemes is $\mathbf{0}_{M\times NL}$. Besides, the dataset $\lbrace \bar{\mathbf{R}}_i,\bar{\mathbf{G}}_i \rbrace ^{K}_{i=1}$ is generated according to (\ref{angular burst model}), (\ref{overall angular burst model}), (\ref{coarse angular real model}) and (\ref{fine angular real model}). The number of samples in the training, validation and test dataset is $K=20000$, $K=5000$ and $K=1000$, respectively. Additionally, training and validation batch sizes are set as 32 and 100, respectively. The learning rate is $0.0005$.  Both the coarse estimation net and the fine correction net use mean square error (MSE) as loss function. The normalized MSE (NMSE) metric is chosen to assess the channel estimation performance, which is defined as follows:
\begin{equation}
{\rm NMSE} = 10log_{10}\bigg(\frac{1}{K}\sum_{i=1}^K \frac{||\bar{\mathbf{G}}_i- \hat{\mathbf{G}}(\mathbf{\Theta},\bar{\mathbf{R}}_{i},\bar{\mathbf{\Phi}} )||_F}{||\bar{\mathbf{G}}_{i}||_F}\bigg)
\end{equation}
where $\mathbf{\Theta}$ stands for both $\lbrace \mathbf{W}^l_{e}, \theta_e^l \rbrace ^{L_{e}}_{l=1}$ and $ \lbrace \mathbf{W}^l_{c}, \mathbf{\theta}_c^l, \omega \rbrace ^{L_{c}}_{l=1}$ in the proposed deep unrolling networks.

Four different schemes that originate from the proposed structure are compared to verify the effectiveness and superiority of our design. The corresponding specified configurations are demonstrated in Table \ref{proposed schemes setting}.
\begin{table*}
\renewcommand{\arraystretch}{1.1}
\caption{Specifications of the proposed schemes. }
\label{proposed schemes setting}
\centering
\begin{tabular}{|m{22mm}|m{22mm}|m{22mm}|m{22mm}|m{22mm}|m{22mm}|m{22mm}|}
\hline
\textbf{Schemes} & \textbf{Coarse Estimation Net} & \textbf{Fine Correction Net} &\textbf{Prior Sparsity Bound} &\textbf{Intra Frame Sparsity} &\textbf{Small-Scale Inter-Frame Sparsity} &\textbf{Large-Scale Inter-Frame Sparsity}\\
\hline
C-F-BSS & $\checked$ & $\checked$ & $\checked$ & $\checked$ & $\checked$  & $\checked$ \\
\hline
C-F-BFSJ & $\checked$ & $\checked$ & $\times$ & $\checked$ & $\checked$  & $\checked$ \\
\hline
C-F-BSS-WS & $\checked$ & $\checked$ & $\checked$ & $\checked$ & $\times$  & $\checked$ \\
\hline
C-F-BFSJ-WS & $\checked$ & $\checked$ & $\times$ & $\checked$ & $\times$  & $\checked$ \\
\hline
F-BSS-WS & $\times$ & $\checked$ & $\checked$ & $\checked$ & $\times$  & $\times$ \\
\hline
F-BFSJ-WS & $\times$ & $\checked$ & $\times$ & $\checked$ & $\times$  & $\times$ \\
\hline
\end{tabular}
\end{table*}

\subsection{Validating the Reasonability of the Proposed schemes}
\begin{figure}
  \centering
  \includegraphics[scale=0.47]{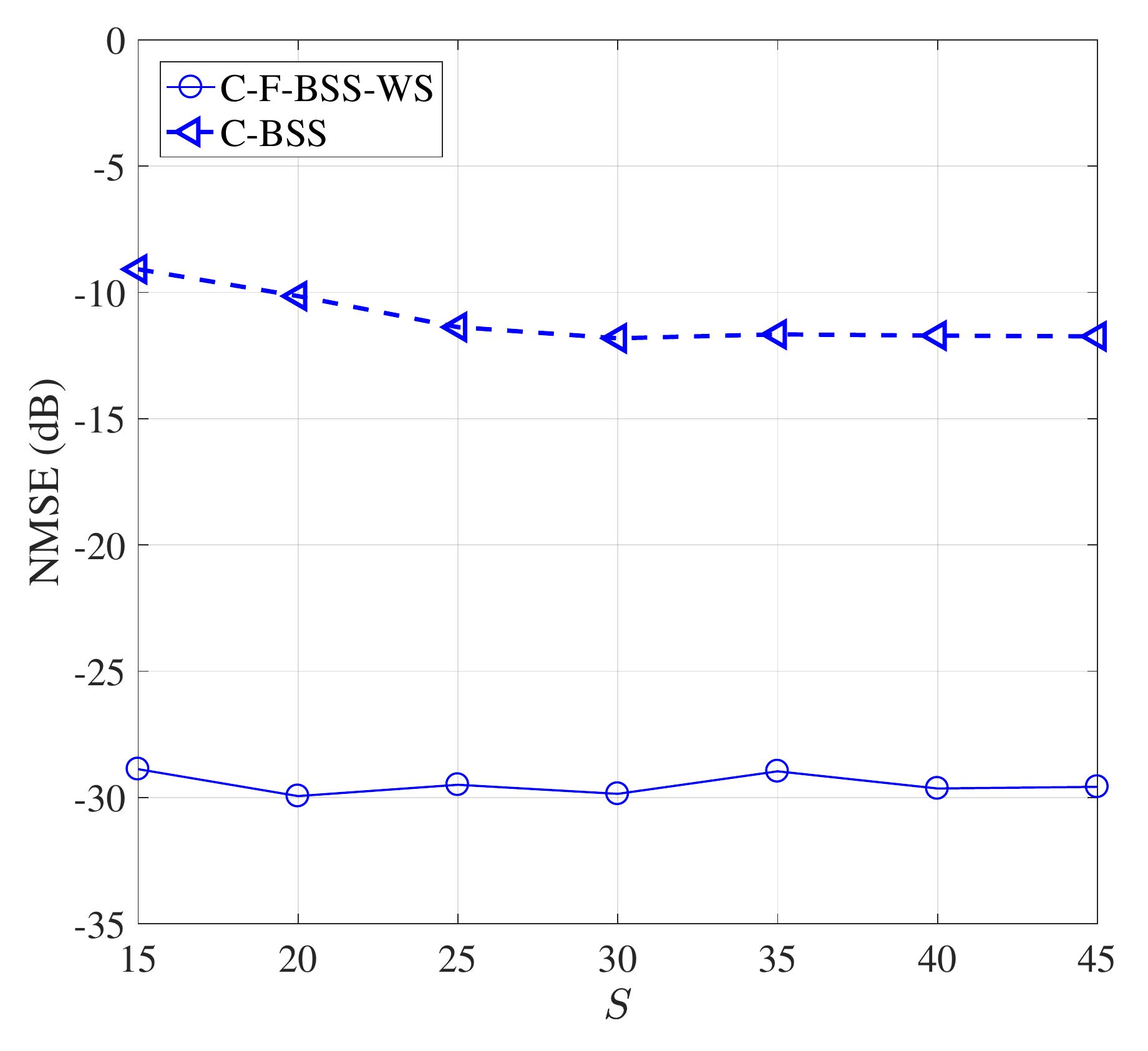}\\
  \caption{NMSE comparison of the coarse estimation net versus the parameter $S$ under $\overline{s}=15$, $T=33$, $s_c=10$, $\text{SNR}=30$ dB.}\label{coarse}
\end{figure}

\begin{figure}
  \centering
  \includegraphics[scale=0.47]{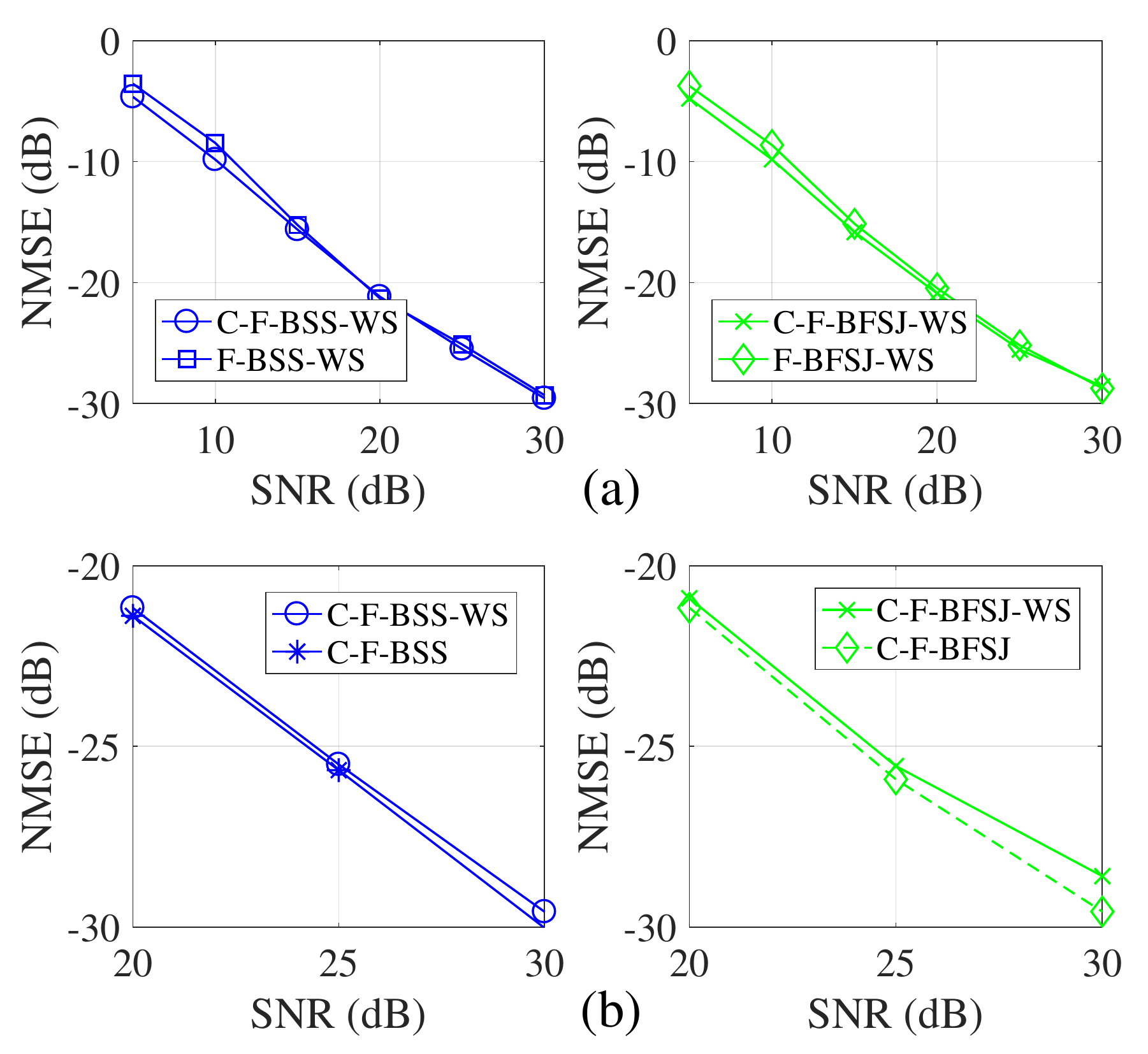}\\
  \caption{NMSE comparison of the proposed networks versus SNR under $\overline{s}=15$, $T=33$ and $s_c=10$.}\label{SNR propose}
\end{figure}

This subsection separately shows the reasonability of each module in our schemes, including the exploitation of large-scale and small-scale inter-frame sparsities and the setup of $S$ in (\ref{coarse S}), which are given in Fig. \ref{coarse}, Fig. \ref{SNR propose}(a) and Fig. \ref{SNR propose}(b), respectively.

In Fig. \ref{coarse}, the impact of the parameter $S$ on C-F-BSS-WS is demonstrated.  Note that under the current setting, i.e., $\overline{s}=15$, $M=128$, $s_c=10$, the parameter $S\approx27$ according to (\ref{coarse S}) by assuming $|\mathbf{\Gamma}^{[1]}|=\frac{s_c+\overline{s}}{2}$. Meanwhile, the performance of C-BSS and C-F-BSS-WS in this figure is close to their best performance when $S$ is set as $27$, which indicates that our proposed formula (\ref{coarse S}) is a proper sparsity bound to assist these two nets in obtaining their best performance.  Moreover, the NMSE performance of C-F-BSS-WS changes slightly when $S$ varies from $15$ to $45$, which means that the result of $S$ calculated based on (\ref{coarse S}) needs not to be strictly accurate and it can be adjusted within a wide range and achieve similar performance.

In Fig. \ref{SNR propose}(a), the proposed C-F-BSS-WS and C-F-BFSJ-WS achieve around $1$ dB gain compared with the F-BSS-WS and F-BFSJ-WS when SNR is not very large, i.e., SNR = $5\sim15$dB.  Note that when $S$ varies from $10$ to $45$, this phenomenon also exists in the proposed schemes, but the detailed comparison is omitted here for simplicity. Based on such observations, we conclude that the coarse estimation net is capable of providing a preliminary channel estimate, which is positive to subsequent fine correction net at the low SNR case. Meanwhile, even without the priori sparsity bound information, the proposed C-F-BFSJ-WS and F-BFSJ-WS have no apparent performance degradation compared with the C-F-BSS-WS and F-BSS-WS networks with varying SNR, which indicates the robustness of the `first significant jump rule'.  

In Fig. \ref{SNR propose}(b), it is demonstrated that the benefit of small-scale inter-frame sparsity appears when SNR is larger than $25$dB. The reason leading to this phenomenon is that the support set $\mathbf{\Gamma}^{[i-1]}$ cannot be reconstructed accurately when SNR is not large enough, so the priori information of $\hat{\mathbf{\Gamma}}^{[i-1]}$ fails to provide obvious benefit in reconstructing the channel in the $i$-th frame.  Therefore, when SNR is relatively small, e.g., SNR = $5\sim15$dB, there is no need to explore small-scale inter-frame sparsity and the NMSE performance of the networks is basically identical whether the small-scale inter-frame sparsity is utilized or not.

\subsection{Channel Estimation Performance of the Proposed Scheme and the Existing Schemes}

\begin{figure}
  \centering
  \includegraphics[scale=0.47]{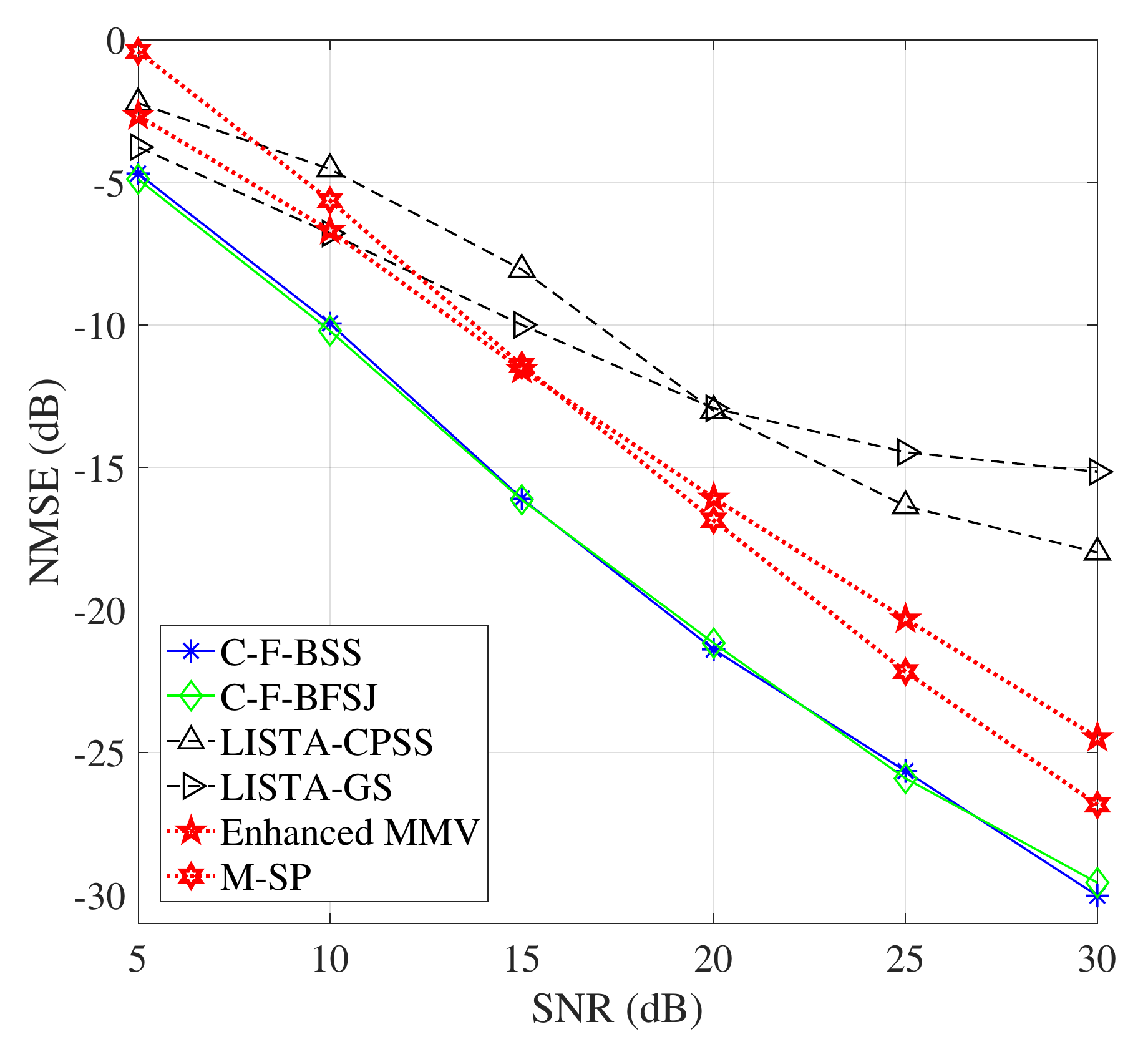}\\
  \caption{NMSE comparison of estimated channel versus SNR under $\overline{s}=15$, $s_c=10$ and $T=33$.}\label{SNR}
\end{figure}

In this section, our proposed schemes are compared with four baseline schemes in Table \ref{existing schemes setting} to demonstrate the superiority of the proposed schemes.
\begin{table*}
\renewcommand{\arraystretch}{1.1}
\caption{Information utilized in the baseline schemes.}
\label{existing schemes setting}
\centering
\begin{tabular}{|m{32mm}|m{32mm}|m{32mm}|m{32mm}|m{32mm}|}
\hline
\textbf{Schemes} & \textbf{Prior Sparsity Bound} &\textbf{Intra-Frame Sparsity} &\textbf{Small-Scale Inter-Frame Sparsity} &\textbf{Large-Scale Inter-Frame Sparsity}\\
\hline
M-SP \cite{9} & $\checked$ & $\checked$ & $\checked$  & $\times$ \\
\hline
Enhanced MMV$\tablefootnote{BCD-MMV algorithm in \cite{21} is utilized to solve the minimization problem in this scheme for fair comparison since our proposed fine correction net also originates from BCD-MMV.}$ \cite{10} & $\checked$ & $\checked$ & $\times$  & $\checked$ \\
\hline
LISTA-CPSS \cite{19} & $\checked$ & $\times$ & $\times$  & $\times$ \\
\hline
LISTA-GS \cite{20} & $\times$ & $\checked$ & $\times$  & $\times$ \\
\hline
\end{tabular}
\end{table*}

In Fig. \ref{SNR}, we compare the NMSE performance of the estimated channel versus SNR. This figure shows that the proposed C-F-BSS and C-F-BFSJ outperform all the baseline schemes with varying SNR, which owes to the intra and inter-frame sparsities that are considered in the network structure. Besides, by combining Fig. \ref{SNR} and Fig. \ref{SNR propose}(a), F-BSS-WS and LISTA-CPSS differ from each other only in the exploitation of intra-frame sparsity but F-BSS-WS performs much better than LISTA-CPSS. Thus we conclude that the exploitation of intra-frame sparsity can greatly improve the NMSE performance of the proposed schemes.

\begin{figure}
  \centering
  \includegraphics[scale=0.47]{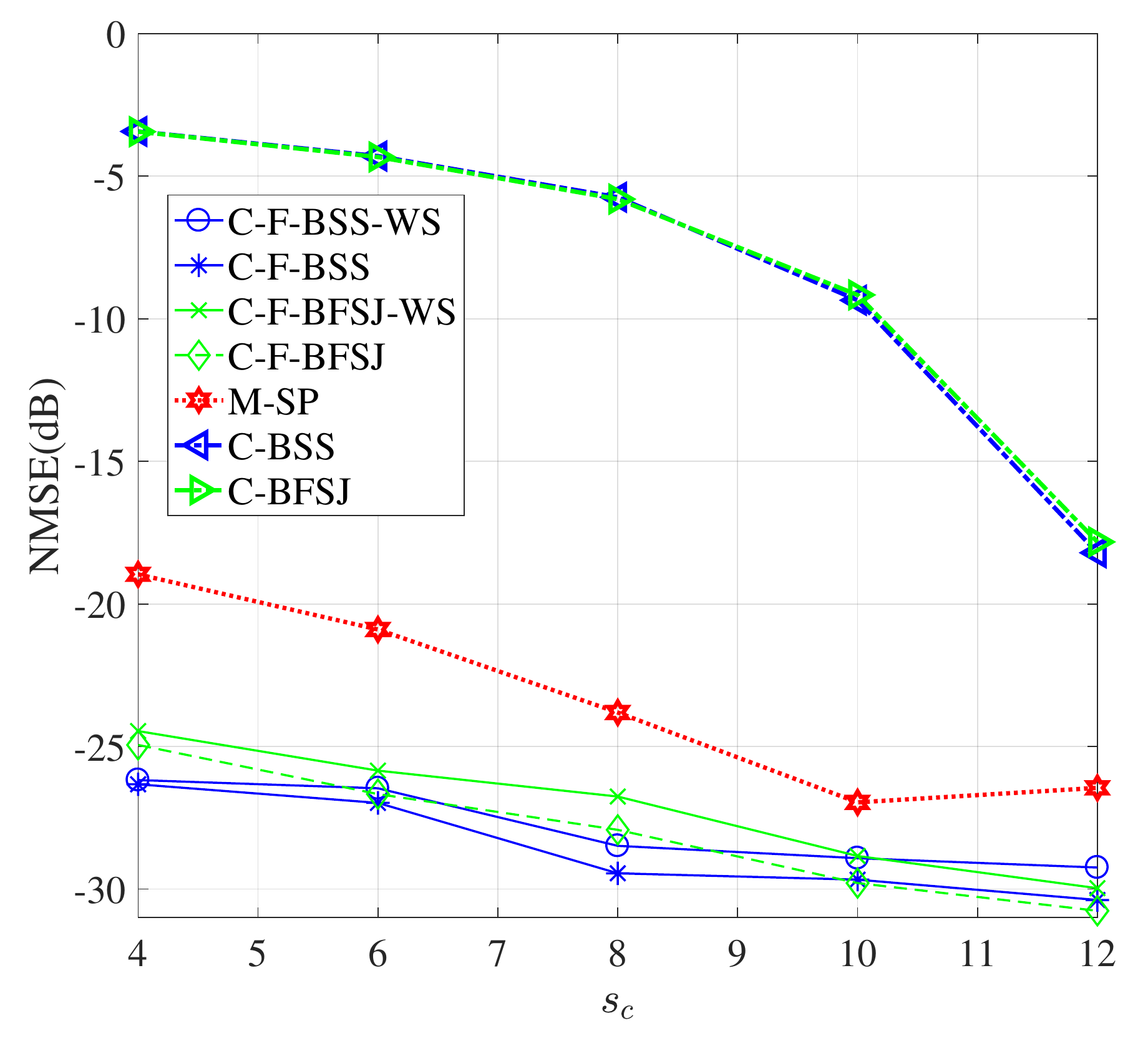}\\
  \caption{NMSE comparison of estimated channel versus the number of shared support indices $s_c$ under $\overline{s}=16$, $S=25$, $T=33$, $|\mathbf{\Gamma}^{[i]}| \sim \mathcal{Z}(\overline{s}-3,\overline{s}-2)$ and $\text{SNR}=30$dB.}\label{shared}
\end{figure}

In Fig. \ref{shared}, NMSE performance of the estimated channel versus the small-scale inter-frame sparsity indicator $s_c$ is depicted. We only compare our schemes with the baseline scheme M-SP because only M-SP considers the small-scale inter-frame sparsity. Meanwhile, we fix $p_l \in [\overline{s}/2M, \overline{s}/M]$ and $S=25$ for different $s_c$ in order to eliminate the impact of these irrelevant parameters. It is clear that our schemes outperform M-SP with varying $s_c$, and the performance gap between schemes using small-scale inter-frame sparsity or not becomes larger as $s_c$ grows, which shows the effectiveness of the utilization of small-scale inter-frame sparsity. The performance of C-F-BSS-WS and C-F-BFSJ-WS is also improved with the increasing $s_c$ even if they do not exploit the small-scale inter-frame sparsity, because the degree of dependence between adjacent frames increases as $s_c$ increases, which results in stronger large-scale inter-frame correlation and consequently improves the performance of the coarse estimation net in the proposed schemes.  Moreover, the performance of C-F-BSS is inferior to C-F-BFSJ when $s_c$ ranges from $10$ to $12$, since we set the $p_{min}$ in (\ref{ss thresholding}) as $\overline{s}/2$ instead of accurate $s_c$.

\begin{figure}
  \centering
  \includegraphics[scale=0.48]{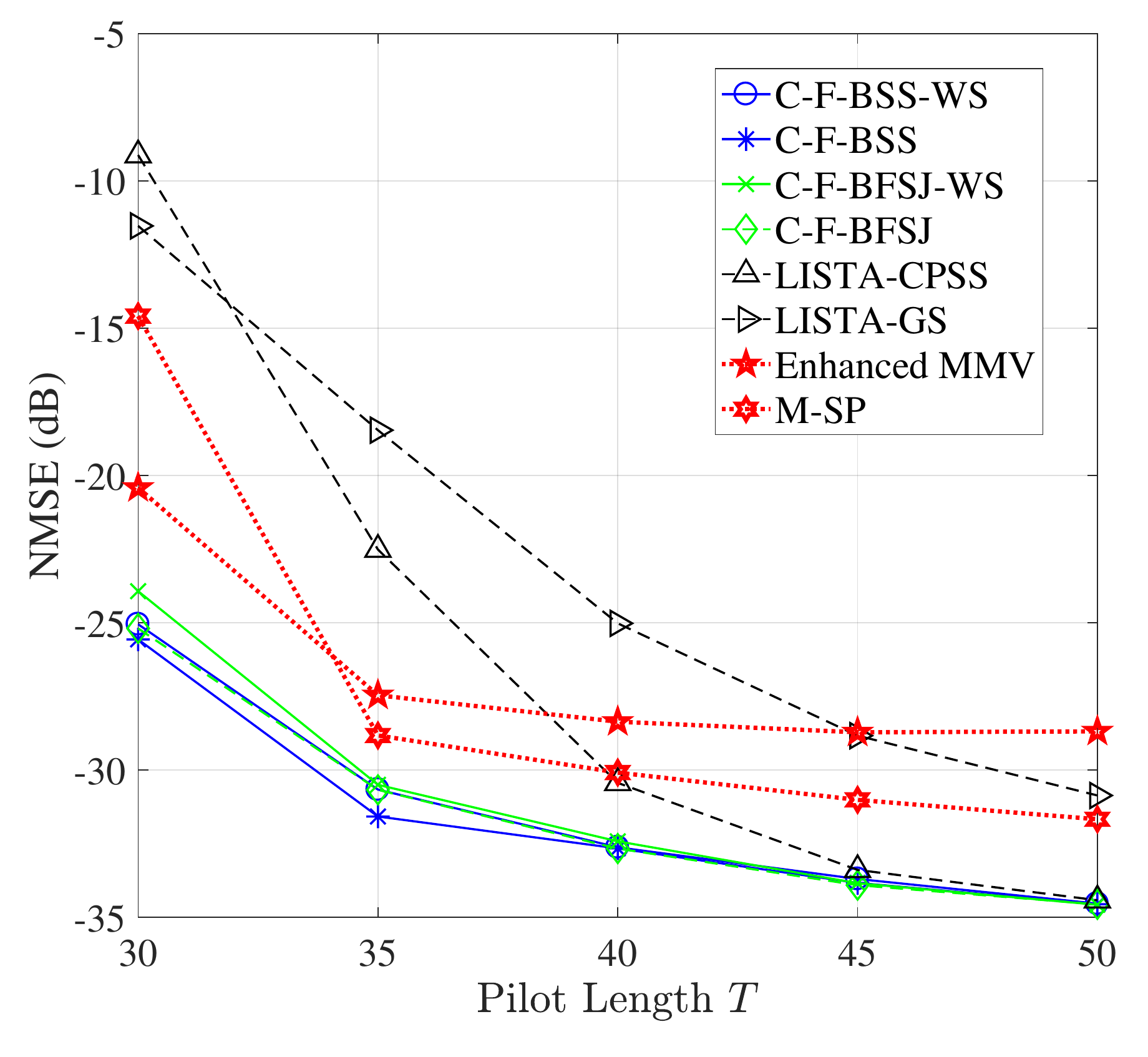}\\
  \caption{NMSE comparison of estimated channel versus pilot length $T$ under $\overline{s}=15$, $s_c=10$ and $\text{SNR}=30$ dB.}\label{CR}
\end{figure}

In Fig. \ref{CR}, it is demonstrated that the performance of the proposed schemes outperforms all the baseline schemes, which implies that fewer pilots are required to achieve identical performance in our schemes. Moreover, when pilot length is relatively small, e.g., $T=30\sim35$, C-F-BSS and C-F-BFSJ achieve $0.6$dB $\sim$ $1.0$dB gain compared with C-F-BSS-WS and C-F-BFSJ-WS, which again confirms the usefulness of the small-scale inter-frame sparsity. However, as pilot length grows larger, the advantage of exploiting small-scale inter-frame sparsity in the proposed schemes gradually vanishes, which implies that there is no need to exploit small-scale inter-frame sparsity when pilot is large enough. 
\begin{table}
\renewcommand{\arraystretch}{1.5}
\caption{Number of multiplications in each iteration to estimate the channel matrix in one frame}
\label{complexity analysis}
\centering
\begin{tabular}{|p{35mm}|p{40mm}|}
\hline
\textbf{Schemes} & \textbf{Computational Complexities}  \\
\hline
Coarse Estimation Net & $8NMT+4NM$  \\
\hline
Fine Correction Net & $8NMT+4NM+4M$ \\
\hline
Enhanced MMV \cite{10} & $8NMT+4NM+4M$ \\
\hline
Modified-SP (M-SP) \cite{9} & $4NMT+3NM+48T\overline{s}^2+24\overline{s}^3+4\overline{s}NT$  \\
\hline
LISTA-GS \cite{20} & $8NMT+4NM+2M$ \\
\hline
LISTA-CPSS \cite{19} & $8NMT$ \\
\hline
\end{tabular}
\end{table}

\subsection{Computational Complexity of Different Schemes}

\begin{figure}
  \centering
  \includegraphics[scale=0.48]{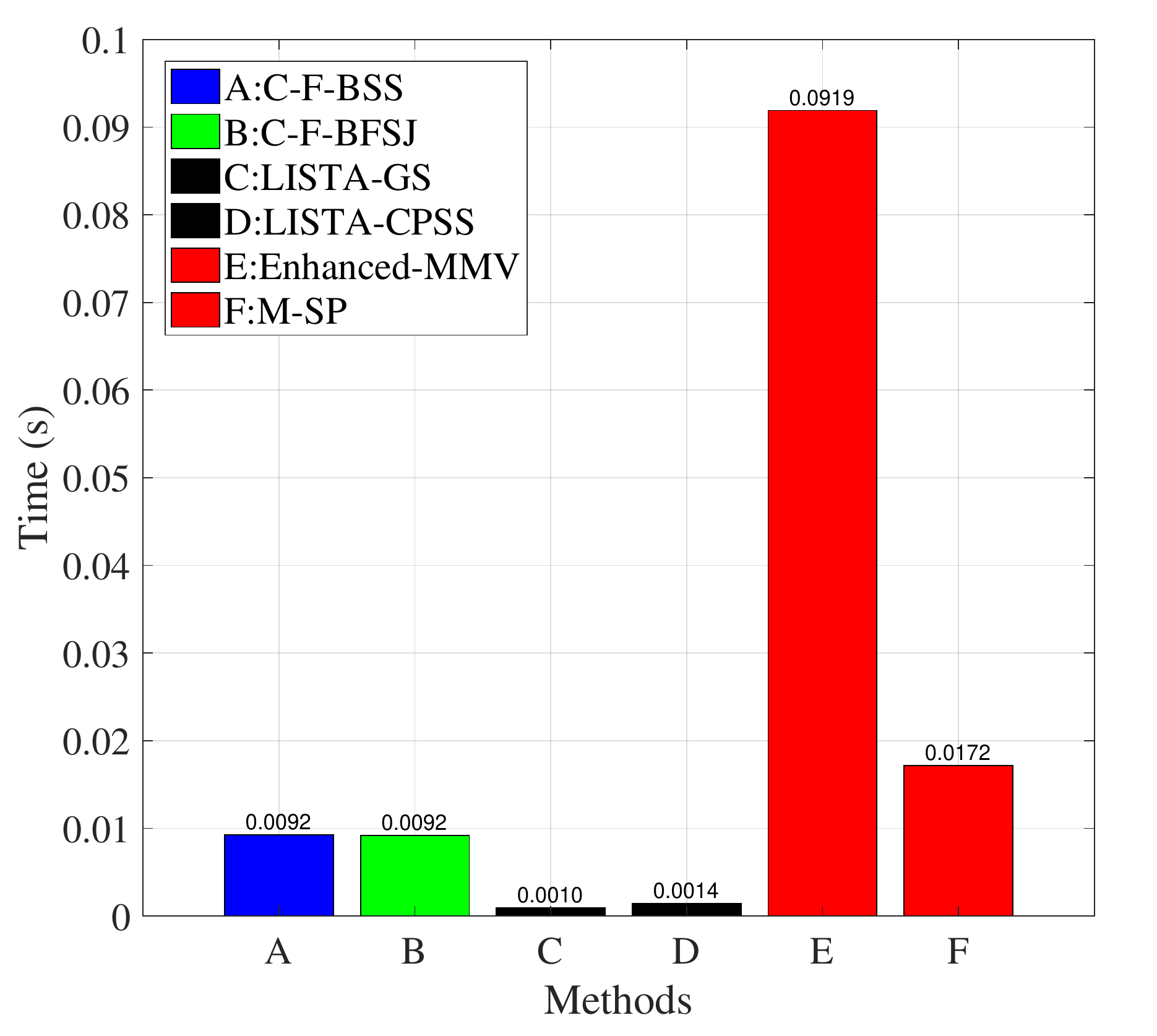}\\
  \caption{Running time comparison of different schemes under $\overline{s}=15$, $T=33$, $s_c=10$ and $\text{SNR}=30$ dB.}\label{time consumption}
\end{figure}

In Fig. \ref{time consumption} and Table \ref{complexity analysis}, the total running time to estimate channel across $L$ frames of all the schemes, and the computational complexity in each iteration of all the schemes are demonstrated, respectively. It is shown in Fig.  \ref{time consumption} that our proposed schemes cost less time than other schemes except for LISTA-GS and LISTA-CPSS. They run faster but at the expense of performance due to the following reasons: First, they do not select proper support set to eliminate shrinkage attenuation or exploit the intra-frame sparsity to calculate $l_2$ norm in (\ref{soft thre support selection function}), respectively. Second, they estimate the channels among $L$ successive frames in parallel. However, our schemes still run faster than other traditional schemes and achieve better trade-off between channel reconstruction performance and time complexity.

\section{Conclusion}
In this paper, we develop a two-stage channel estimation structure in downlink FDD MIMO system, which consists of a coarse estimation net and a fine correction net. The former net regards the received frames as a unity and exploits the large-scale inter-frame sparsity in terms of row sparsity via $l_{2,1}$ minimization. The latter net is designed to further refine the estimated channel coefficients of the former net in a frame-by-frame manner, which exploits the small-scale inter-frame sparsity between adjacent frames through weighted $l_{2,1}$ minimization. Moreover, two schemes are proposed depending on whether prior sparsity information is required. The first one is C-F-BSS consisting of C-BSS and F-BSS, which extends SS thresholding function in \cite{19} to BSS to relieve the magnitude attenuation but requires priori sparsity bound information. The parameter that determines the selected support set cardinality in C-BSS and its detailed analysis are given. The second one is C-F-BFSJ consisting of C-BFSJ and F-BFSJ, which generalizes BSS based on the `first significant jump' rule to eliminate the requirement of priori sparsity bound information without drastically degrading performance.

Simulations results show that the proposed schemes can accurately estimate the channel with fewer pilot overhead and lower SNR compared with all baseline schemes, and the complexity of the proposed schemes is lower than that of most baseline schemes except LISTA-GS and LISTA-CPSS which require more pilot overhead. Meanwhile, the requirement of priori sparsity bound information is eliminated in the proposed scheme with desirable performance.

\appendices

\section{THE PROOF CONCERNING THE AVERAGE NUMBER OF THE NONZERO ROWS IN (\ref{coarse S})}

Firstly, when $\mathbf{G}$ contains channels across $L-1$ frames, we assume that the exact number of nonzero rows in $\mathbf{G}$ is $m_{L-1} = |\cup_{i=1}^{L-1} \mathbf{\Gamma}^{[i]}|$ and the exact number of nonzero rows in $\mathbf{S}^{[L-1]}$ is $n_{L-1}=|\mathbf{\Gamma}^{[L-1]}|$.  As the channel $\mathbf{S}^{[L]}$ is concatenated with $\mathbf{G}$, the indices in its support set $\mathbf{\Gamma}^{[L]}$ can be classified into three kinds: 

1. The indices that overlap with the set $\mathbf{\Gamma}^{[L-1]}$.

2. The indices that are included in $\cup_{i=1}^{L-1} \mathbf{\Gamma}^{[i]} \backslash \mathbf{\Gamma}^{[L-1]}$. 

3. The indices that are included in $\lbrace 1,\cdots,M \rbrace \backslash \cup_{i=1}^{L-1} \mathbf{\Gamma}^{[i]}$.

\begin{figure}[h]
  \centering
  \includegraphics[scale=0.36]{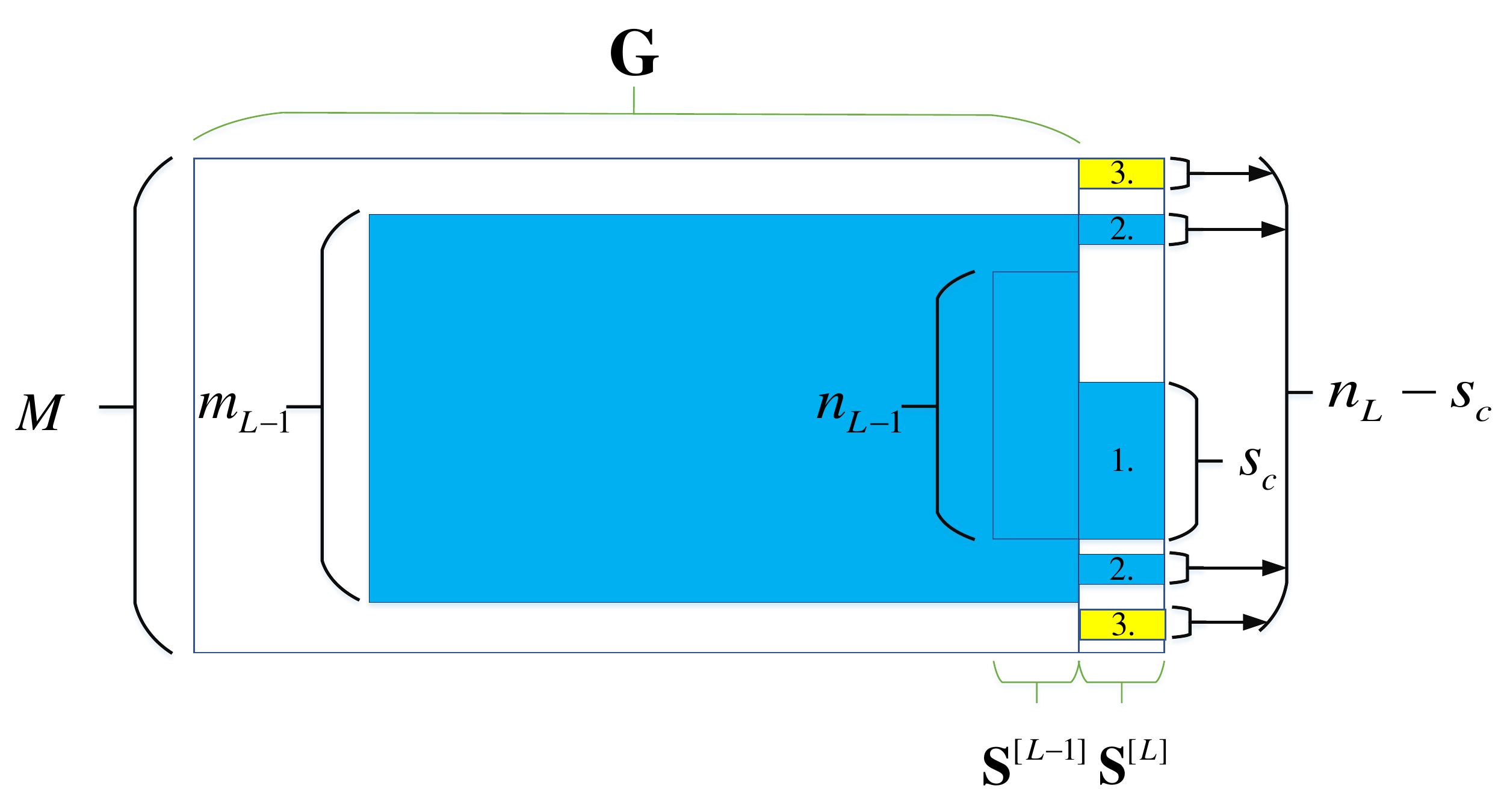}\\
  \caption{Explanation of the relationship of the indices.}\label{proof}
\end{figure}

For a better understanding of the above relationship, a simple illustration of the channel matrix $\mathbf{G}$ with $M$ rows is given in Fig. \ref{proof}. Its nonzero rows occupy $m_{L-1}$ rows of the $M$ rows, which are shown in blue. After the channel $\mathbf{S}^{[L]}$ is concatenated with $\mathbf{G}$, its indices of the nonzero rows that correspond to different kinds mentioned above are marked with the corresponding number.

Furthermore, we know that $s_c$ is the lower bound of the number of indices classified into the first kind, so that the upper bound of the total number of indices that are classified into the second kind and the third kind is $n_{L}-s_c$. However, whether the $j$-th index ($j \in \lbrace 1,\cdots,n_{L}-s_c \rbrace$) belongs to the second kind or the third kind is random, which leads to the random number of the indices of either kind. Meanwhile, by following the assumption that the angles of departure are uniformly distributed over $(0,\pi)$, we assume that signal departs from each angle affected by the individual scatterers with equal probability (the indices belonging to the last two kinds represent the angles affected by the individual scatterers). Therefore, if we choose the $j$-th index from $n_{L}-s_c$ indices of the last two kinds, whether the $j$-th index belongs to the second kind or the third kind follows the bernoulli distribution: 
\begin{equation}
\begin{aligned}
\begin{cases}
p_{nt} = \frac{m_{L-1}-n_{L-1}}{M-n_{L-1}}, & j\ \text{in the second kind}, \\
p_{t} = \frac{M-m_{L-1}}{M-n_{L-1}}, & j\ \text{in the third kind},
\end{cases}
\end{aligned}
\end{equation}
where $p_{nt}$ and $p_{t}$ represent the probability that the $j$-th index belongs to the second kind and the third kind, respectively.

Therefore, the probability that there are $k$ indices out of $n_{L}-s_c$ indices classified as the third kind is defined as follows:
\begin{equation}
    P\lbrace X=k \rbrace = C^k_{n_{L}-s_c} {p}^k_t(1-p_t)^{n_{L}-s_c-k},
\end{equation}
and the corresponding distribution is binomial distribution which has an expectation of $(n_{L}-s_c)p_{t}$. 

Based on the above arguments, when the channel in the $L$-th frame is concatenated with $\mathbf{G}$, its average number of nonzero rows $m_{L}$ is given as:

\begin{equation}\label{expectation non zero row}
\begin{aligned}
\mathbbm{E}(m_{L}) & = \mathbbm{E}((n_{L} - s_c)p_t) + \mathbbm{E}(m_{L-1}), \\
& = (\mathbbm{E}(n_{L} - s_c))\frac{M-\mathbbm{E}(m_{L-1})}{M-\mathbbm{E}(n_{L-1})} + \mathbbm{E}(m_{L-1}), \\
& = \frac{(\mathbbm{E}(n_{L} - s_c))M}{M-n_{L-1}} + \frac{\mathbbm{E}(m_{L-1})(M-2\mathbbm{E}(n_{L-1})+s_c)}{M-n_{L-1}}.
\end{aligned}
\end{equation}
Note that we assume: 
\begin{equation}
\begin{aligned}
&\mathbbm{E}(n_{1})= \cdots = \mathbbm{E}(n_{L}) = \mathbbm{E}(|\mathbf{\Gamma}^{[1]}|) = \cdots =\mathbbm{E}(|\mathbf{\Gamma}^{[L]}|) = \frac{\overline{s}+s_c}{2}, \\
& a = \frac{M(\mathbbm{E}(n_{L}) - s_c)}{M-n_{L-1}}, \\
& b = \frac{M-\mathbbm{E}(n_{L-1})-\mathbbm{E}(n_{L})+s_c}{M-n_{L-1}}.
\end{aligned}
\end{equation}
The equation (\ref{expectation non zero row}) is now transformed with $\mathbbm{E}(m_{1}) = \mathbbm{E}(n_{1}) = \mathbbm{E}(|\mathbf{\Gamma}^{[1]}|)$:
\begin{equation}
\begin{aligned}
\mathbbm{E}(m_{L}) & = a + b\mathbbm{E}(m_{L-1}), \\
& = b^{L-1}\mathbbm{E}(m_{1}) + \frac{a(1-b^{L-1})}{1-b},
\end{aligned}
\end{equation}
Finally, as $\mathbf{G}$ and $\bar{\mathbf{G}}$ follows the following equation:
\begin{equation}
\bar{\mathbf{G}} = \begin{bmatrix} \mathbbm{R}(\mathbf{G})  \\ \mathbbm{I}(\mathbf{G}) \end{bmatrix},
\end{equation}
we can derive the result in (\ref{coarse S}):
\begin{equation}
S = 2\mathbbm{E}(m_{L}).
\end{equation}
\section{COMPLEXITY ANALYSIS IN TABLE \ref{complexity analysis}}
The computational complexities of the coarse estimation net and the fine correction net in each iteration are analyzed as follows. Here we only consider steps that contribute to most of the system computation complexity, i.e., the step 3 in \textbf{Network} \ref{alg:coarse estimation net} and the step 5 in \textbf{Network} \ref{alg:fine correction net}.

Complexity analysis for step 3 of the coarse estimation net:  The complexity of the matrix multiplication in step 3 is $O(8NMTL)$. Meanwhile,  the $l_2$ norm operation and the product of the thresholding function and the corresponding row in (\ref{soft thre support selection function}) or (\ref{soft thre first signifdicant jump function})  have the same computational complexity $O(2NML)$. Therefore the number of multiplications in each iteration is $O(8NMTL+4NML)$. As a result, the number of multiplications in each iteration for one frame is $O(8NMT+4NM)$ for both C-BSS and C-BFSJ. 

Complexity analysis for step 5 of the fine correction net:  The complexity of the matrix multiplication in step 5 is $O(8NMT)$. Similarly, the computational complexity of the $l_2$ norm operation is $O(2MN)$. The computational complexity of the product between the thresholding $\eta_{\theta_c^l\bm{\omega}}(\cdot)$ and the corresponding row in (\ref{soft weighted thre support selection function}) is $O(2MN+2M)$. The complexity of the product between $\theta_c^l$ and $\bm{\omega}$ is $O(2M)$. Therefore, the number of multiplications in each iteration is $O(8NMT+4NM+4M)$ for both F-BSS and F-BFSJ.

\section*{Acknowledgment}

This work was supported by the National Natural Science Foundation of China (61871050), U.S. National Science Foundation (2136202).

Thanks Yandong Shi from ShanghaiTech University Shanghai, China for providing his code.

\bibliographystyle{ieeetran}
\bibliography{first_jrnl}

\end{document}